\documentclass[12pt,english]{article}
\usepackage{times}
\usepackage[T1]{fontenc}
\usepackage[latin1]{inputenc}
\usepackage{geometry}
\usepackage{array}
\usepackage{subfigure}
\usepackage{graphicx}
\usepackage{setspace}
\makeatletter

\providecommand{\tabularnewline}{\\}

 \newcommand{\lyxaddress}[1]{
   \par {\raggedright #1 
   \vspace{1.4em}
   \noindent\par}
 }

\usepackage{array}
\usepackage{subfigure}
\usepackage{graphicx}
\usepackage{setspace}

\usepackage{setspace}

\usepackage{setspace}

\usepackage{setspace}

\newcommand{\ovec}[1]{{\mbox{\boldmath $#1$}}}

\newcommand{\bB}{\mathrm{\mathbf{B}}}

\newcommand{\be}{\ovec{e}}

\newcommand{\bu}{\mathrm{\mathbf{u}}}

\newcommand{\bzeta}{\ovec{\zeta}}

\newcommand{\bmB}{\overline{\mathrm{\mathbf{B}}}}
\newcommand{\mB}{\overline{B}}
\newcommand{\bscE}{\ovec{\cal{E}}}

\newcommand{\bmu}{\overline{\mathrm{\mathbf{u}}}}

\newcommand{\bnab}{\nabla}
\def\dd {\mbox{d}}

\def\iu {\mbox{i}}
\def\x {\times}

\def\p {\partial}
\def\bzo {\bf 0}

\usepackage{babel}
\makeatother
\begin{document}
\begin{description}
\item [Authors:]~
\end{description}
\begin{singlespace}
\noindent RAUL ALEJANDRO AVALOS-ZU\~{N}IGA
\end{singlespace}

\begin{onehalfspace}
\noindent Universidad Aut\'{o}noma Metropolitana-Iztapalapa. Av.
San Rafael Atlixco 186, col. Vicentina, 09340 D.F. M\'{e}xico. Tel.: +52
55 5804 4648 ext. 238. Fax: +52 55 5804 4900. E-mail: raaz@xanum.uam.mx. 
\medskip{}
\end{onehalfspace}

\begin{onehalfspace}
\medskip{}
\noindent MINGTIAN XU
\end{onehalfspace}

\begin{onehalfspace}
\noindent Forschungszentrum Dresden-Rossendorf, P.O. Box 510119, 01314
Dresden, Germany. Tel.: +49 351 260 2227. Fax: +49 351 260 2007. E-mail:M.Xu@fzd.de
\medskip{}
\end{onehalfspace}

\begin{onehalfspace}
\medskip{}
\noindent FRANK STEFANI
\end{onehalfspace}

\begin{onehalfspace}
\noindent Forschungszentrum Dresden-Rossendorf, P.O. Box 510119, 01314
Dresden, Germany. Tel.: +49 351 260 3069. Fax: +49 351 260 2007. E-mail:
F.Stefani@fzd.de
\medskip{}
\end{onehalfspace}

\begin{onehalfspace}
\medskip{}
\noindent GUNTER GERBETH
\end{onehalfspace}

\begin{onehalfspace}
\noindent Forschungszentrum Dresden-Rossendorf, P.O. Box 510119, 01314
Dresden, Germany. Tel.: +49 351 260 2168. Fax: +49 351 260 2007. E-mail:G.Gerbeth@fzd.de
\medskip{}
\end{onehalfspace}

\begin{onehalfspace}
\medskip{}
\noindent FRANCK PLUNIAN
\end{onehalfspace}

\begin{onehalfspace}
\noindent Laboratoire de G\'{e}ophysique Interne et de Tectonophysique,
BP 53, 38041 Grenoble Cedex 9, France. Tel.: +33 4 76 82 80 37. Fax: 33 4 76 82 81 01. E-mail: Franck.Plunian@ujf-grenoble.fr
\end{onehalfspace}

\title{\textbf{\large Cylindrical anisotropic $\alpha^{2}$ dynamos}}

\date{R. AVALOS-ZU\~{N}IGA%
\footnote{Email: raaz@xanum.uam.mx. Current Address: Universidad Aut\'{o}noma
Metropolitana-Iztapalapa. Av. San Rafael Atlixco 186, col. Vicentina,
09340, D.F., M\'{e}xico.%
} $\dagger$, M. XU$\dagger$, F. STEFANI$\dagger$, G. GERBETH$\dagger$\\
 and F. PLUNIAN$\ddagger$}

\maketitle

\lyxaddress{\begin{center}$\dagger${\small Forschungszentrum Dresden-Rossendorf,
P.O. Box 510119, 01314 Dresden, Germany}\\
 $\ddagger${\small Lab. de G\'{e}ophysique Interne et de Tectonophysique,
BP 53, 38041 Grenoble Cedex 9, France}\\
(Received 30 November 2006; in final form 15 June 2007)\end{center}}

{\footnotesize We explore the influence of geometry variations on
the structure and the time-dependence of the magnetic field that is
induced by kinematic $\alpha^{2}$ dynamos in a finite cylinder. The
dynamo action is due to an anisotropic $\alpha$ effect which can
be derived from an underlying columnar flow. The investigated geometry
variations concern, in particular, the aspect ratio of height to radius
of the cylinder, and the thickness of the annular space to which the
columnar flow is restricted. Motivated by the quest for laboratory
dynamos which exhibit Earth-like features, we start with modifications
of the Karlsruhe dynamo facility. Its dynamo action is reasonably
described by an $\alpha^{2}$ mechanism with anisotropic $\alpha$
tensor. We find a critical aspect ratio below which the dominant magnetic
field structure changes from an equatorial dipole to an axial dipole.
Similar results are found for $\alpha^{2}$ dynamos working in an
annular space when a radial dependence of $\alpha$ is assumed. Finally,
we study the effect of varying aspect ratios of dynamos with an $\alpha$
tensor depending both on radial and axial coordinates. In this case
only dominant equatorial dipoles are found and most of the solutions
are oscillatory, contrary to all previous cases where the resulting
fields are steady.}{\footnotesize \par}

\noindent \textit{\footnotesize Keywords:} {\footnotesize Dynamo;
$\alpha$ effect; magnetic field orientation}{\footnotesize \par}

\section{Introduction}

It is generally assumed that columnar flows in the Earth's outer core
play an essential role for the generation of the geomagnetic field.
At the surface of the Earth, the magnetic field structure has almost
an axial dipole (AD) structure closely aligned with the Earth's rotation
axis. Direct numerical simulations of the geodynamo have successfully
reproduced many observed features like, e.g., the dominance of the
axial dipole and the occurrence of reversals (e.g. Olson \textit{et
al.} 1999, Ishihara and Kida 2002, Aubert and Wicht 2004, Wicht and
Olson 2004 and references therein). The poloidal part of the field
is thought to be produced from the toroidal part by the $\alpha$-effect
generated by the columnar flows, while the toroidal component of the
Earth's magnetic field is associated to the $\Omega$-effect, but
also again to an $\alpha$-effect or even to both mechanisms together.
These types of magnetic field generation are usually referred to as
$\alpha\Omega$, $\alpha^{2}$ and $\alpha^{2}\Omega$ dynamos, respectively.

It was one of the motivations of the Karlsruhe dynamo experiment to
study an Earth-like magnetic field generation process in the laboratory
(Gailitis 1967, Busse 1975, Stieglitz and M\"{u}ller 2001). However,
in contrast to the axial dipole (AD) of the Earth, the eigenfield
structure of the Karlsruhe dynamo is an equatorial dipole (ED) what
has been predicted in terms of the mean-field theory with an anisotropic
$\alpha$ effect (R\"{a}dler \textit{et al.} 1998). Actually, a general
tendency of anisotropic $\alpha^{2}$ dynamos to produce fields with
dominant equatorial dipole structure has been known for long (R\"{a}dler
1975, R\"{a}dler 1980, R\"{u}diger 1980, R\"{u}diger and Elstner
1994).

It is also well known that a transition from equatorial to axial dipoles
can occur if some differential rotation is added (R\"{a}dler 1986,
Gubbins \textit{et al.} 2000). However, an axial field orientation
can also result from $\alpha^{2}$ dynamos if the magnetic diffusion
is enhanced by small scales of the flow (Tilgner 2004).

Besides the axial and equatorial dipole, the quadrupole structure
seems to play also a certain role in geodynamo models. In many kinematic
models one finds a quasi-degeneration with the dipole field (Gubbins
\textit{et al.} 2000). This degeneration is also responsible for the
appearance of hemispherical dynamos in dynamically coupled models
(Grote and Busse 2000). In this case, both quadrupolar and dipolar
components contribute nearly equal magnetic energy so that their contributions
cancel in one hemisphere and add to each other in the opposite hemisphere.
The interplay between the nearly degenerated (Gubbins \textit{et al.}
2000) axial dipole, equatorial dipole, and quadrupole was used in
various models to explain the reversal phenomenon of the geodynamo
(Melbourne \textit{et al.} 2001).

With the same focus on field reversals, the importance of transitions
between steady and oscillatory solutions of kinematic dynamos has
been highlighted by several authors (Weisshaar 1982, Yoshimura \textit{et
al.} 1984, Sarson and Jones 1999, Phillips 1993, R\"{u}diger \textit{et
al.} 2003). In an extremely reduced reversal model dealing only with
the axial dipole it was shown that many features of reversals (typical
time scales, asymmetry between slow dipole decay and fast recovery,
bimodal field distribution) can be understood by the magnetic field
dynamics in the vicinity of transition points between steady and oscillatory
solutions (Stefani and Gerbeth 2005, Stefani \textit{et al.} 2006a,
Stefani \textit{et al.} 2006b). The main ingredient of this reversal
model, as well as of the reversal model of Giesecke et al. (Giesecke
\textit{et al.} 2005a), is a sign change of $\alpha$ along the radius
which brings into play a coupling between the first two radial eigenfunctions
of the axial dipole field. It should be noticed that such a sign change
results indeed from simulations of magneto-convection (Giesecke \textit{et
al.} 2005b).

With this background, we investigate in the present paper various
kinematic dynamo models within cylindrical geometry. Our focus will
lay first on the dominance of field structure: equatorial (ED) or
axial dipoles (AD) or even quadrupoles (Q), and second on the occurrence
of oscillatory solutions. The cylindrical geometry, which might seem
awkward from the purely geodynamo perspective, is quite natural from
an experimentalist's viewpoint. One could ask, e.g., how the geometry
and the arrangement of spin-generators in the Karlsruhe dynamo could
be modified in order to make its eigenfield prone to reversals.

After presenting the general framework, we will explore geometrical
effects that could lead to dominant AD fields in cylindrical anisotropic
$\alpha^{2}$ dynamos. The utilised numerical code, which is based
on the integral equation approach to kinematic dynamos (Stefani \textit{et
al.} 2000, Xu \textit{et al.} 2004a, Xu \textit{et al.} 2004b, Xu
\textit{et al.} 2006), was already used for the simulation of various
cylindrical dynamos, including the VKS dynamo experiment in Cadarache
(Stefani \textit{et al.} 2006c).

The geometrical variations which are actually considered are the aspect
ratio of height to radius of the cylinder and the width of the annular
space to which the dynamo source is restricted. First we consider
the geometry of the Karlsruhe dynamo experiment. Its steady dynamo
field is generated by a bundle of axially invariant helical columns,
which is well described within mean-field theory as an $\alpha^{2}$
dynamo with anisotropic $\alpha$-effect. We find that for this experiment
a dominant AD field could be achieved below a critical value of the
aspect ratio which is not so far from the one of the real facility.
In a next step, we explore more complex structures of $\alpha$ which
have been derived from a flow described by axially invariant helical
columns which are restricted to an annular space. The resulting $\alpha$
coefficients acquire a radial profile which depends on the flow structure.
As for the modified Karlsruhe case, dynamo solutions show dominant
steady AD fields below a critical value of the aspect ratio. In contrast
to this, the reduction of the thickness of the annular space does
not lead to a transition from non-axisymmetric to axisymmetric modes,
although the critical dynamo numbers for both modes seem to converge.
Finally, we have considered axial-radial dependence of $\alpha$.
The dynamo action works in a fixed annular space and again the aspect
ratio of height to radius of cylinder is the varying geometrical parameter.
In this case, non-axisymmetric oscillatory fields are the dominant
solutions.

\section{The general concept}

We consider an incompressible steadily moving fluid with velocity
$\bu$, which is confined to a cylinder and surrounded by vacuum.
The fluid has homogenous electrical conductivity $\sigma$ and magnetic
permeability $\mu$. The fluid motion induces a magnetic field $\bB$
which extends in whole space. The magnetic field is governed by the
induction equation\begin{equation}
\eta\bnab^{2}\,\bB+\bnab\x(\bu\x\bB)-\p_{t}\,\bB=\bzo\,,\quad\bnab\cdot\bB=0\,,\label{eq:induction}\end{equation}
 where $\eta$ is the magnetic diffusivity defined by $\eta=1/\mu\sigma$.
In the mean field approach, each quantity is decomposed into a mean
part (denoted by an overline) and a fluctuating part (denoted by a
prime). Referring to a cylindrical coordinate system $(s,\varphi,z)$,
we define mean fields by averaging over $\varphi$.

As we are only interested in the induction effects originated by the
fluctuating part $\bu'$ we assume that the mean motion $\bmu$ is
equal to zero. In this case, the mean part of the induction equation
(1) reduces to \begin{equation}
\eta\bnab^{2}\,\bmB+\bnab\x\bscE-\p_{t}\,\bmB=\bzo\,,\quad\bnab\cdot\bmB=0\,,\label{eq:mean-ind}\end{equation}
 where $\bscE=\overline{\bu'\x\bB'}$, is the mean electromotive force
(e.m.f.) which is the source of generation of the large scale magnetic
field $\bmB$. This e.m.f. results from the interaction of motion
and magnetic field at small scales.

\subsection{Representations of $\bscE$}

We consider different forms of $\bscE$ which are generated by flows
organised in columnar vortices parallel to the vertical axis of the
cylinder. In the following only the $\alpha$ effect that results
from these columnar structures is considered as the main contribution
to the generation of $\bscE$, other effects are just neglected. In
a strict sense, such a reduction of $\bscE$ to an $\alpha$-effect
term is only possible if the spatial variations of $\bmB$ are sufficiently
weak.

We consider first the mean e.m.f $\bscE$ produced by the flow in
the Karlsruhe dynamo experiment (R\"{a}dler \textit{et al.} 1998
). Its most simplified analytical representation is given by \begin{equation}
\begin{array}{ccc}
\bscE & = & -\alpha\left(\,\bmB-(\mathrm{\mathbf{e}}_{z}\cdot\bmB\,)\mathrm{\mathbf{e}}_{z}\right).\end{array}\label{eq:efm-Karlsruhe}\end{equation}
 where $\alpha$ is constant in the cylindrical volume and ${\be}_{z}$
is the unit vector in axial direction. We point out the anisotropy
of the $\alpha$-effect as represented in (\ref{eq:efm-Karlsruhe}).

The next considered example of $\bscE$ results from an axially invariant
flow organised in columnar vortices equally distributed in an annular
region. Similar flow structures were recently discussed in the context
of quasi-geostrophic dynamos (Schaeffer and Cardin 2006). A detailed
description of such ''rings of rolls'' and the $\alpha$ tensor
resulting from them has been derived by Avalos et al. (2007) and is
given in Appendix A. The mean e.m.f. $\bscE$ produced by such a flow
is given, under the assumptions mentioned in Appendices A and B, by
\begin{equation}
{\mathcal{{E}}}_{\kappa}=\alpha_{\kappa\lambda}(s)\,\mB_{\lambda},\label{eq:fem-zind}\end{equation}
 with the subscripts $\kappa$ and $\lambda$ standing for $s$, $\varphi$,
or $z$.

For some flow configurations, it has been shown (Avalos \textit{et
al.} 2007) that the resulting matrix $\alpha_{\kappa\lambda}$ is
of the form \begin{equation}
\alpha_{\kappa\lambda}=\left\{ \begin{array}{ccc}
\alpha_{ss}(s) & 0 & 0\\
0 & \alpha_{\varphi\varphi}(s) & 0\\
0 & \alpha_{z\varphi}(s) & 0\end{array}\right\} .\label{eq:matrix-alpha}\end{equation}

We have also considered an additional axial dependence of the components
in (\ref{eq:matrix-alpha}) multiplying them by harmonic functions
of $z$ that vanish at the top and the bottom of the cylinder. This
was motivated by the fact that for rolls in real rotating bodies a
North-South antisymmetry of the axial velocity is expected, while
the horizontal velocity components are expected to be symmetric with
respect to the equator. Admittedly, the correct treatment of this
problem would require a new derivation of the $\alpha$ matrix for
such rolls along the lines outlined in the appendices. As a sort of
compromise we focus here only on the general symmetry properties of
the elements of the $\alpha$ matrix. Since $\alpha_{ss}$ and $\alpha_{\varphi\varphi}$
depend on products of axial and horizontal velocity components, we
expect an antisymmetric behaviour. On the other hand, $\alpha_{z\varphi}$
should remain North-South symmetric since it depends on horizontal
velocity components only.

The cylinder is assumed to extend over the axial interval $-H/2\leq z\leq H/2$.
If we assume $u_{s}$ and $u_{\varphi}$ to be proportional to $\cos(\pi z/H)$
and $u_{z}$ to be proportional to $\sin(2\pi z/H)$, then the new
$\alpha$ matrix is given by \begin{equation}
\alpha_{\kappa\lambda}=\left\{ \begin{array}{ccc}
\alpha_{ss}(s)\cos(\pi z/H)\sin(2\pi z/H) & 0 & 0\\
0 & \alpha_{\varphi\varphi}(s)\cos(\pi z/H)\sin(2\pi z/H) & 0\\
0 & \alpha_{z\varphi}(s)\cos^{2}(\pi z/H) & 0\end{array}\right\} .\label{eq:matrix-alpha-z}\end{equation}
 Evidently, the resulting diagonal elements of $\alpha_{\kappa\lambda}$
are anti-symmetric with respect to $z=0$, whereas the non-diagonal
element is symmetric with respect to $z=0$.

Though the representation of $\bscE$ defined above was derived for
an infinitely extended conducting fluid, we assume that it applies
also to a finite cylinder. This approximation has been successfully
used, e.g. in R\"{a}dler \textit{et al.} (1998, 2002), to solve the Karlsruhe
dynamo problem in a finite cylinder. In this case the symmetry of
the most easily excited magnetic field mode was found to be independent
of the conductivity outside the cylinder, while the other properties
of this mode well depend on the conductivity in outer space.

\section{Dynamo solutions}

Once we have defined different representations of $\bscE$, we solve
the mean-field dynamo problem in a finite cylinder enclosed by vacuum
using a numerical code based on the integral equation approach (Stefani
\textit{et al.} 2000, Xu \textit{et al.} 2004a, Xu \textit{et al.}
2004b, Xu \textit{et al.} 2006).The magnetic field is determined by
a self-consistent solution of the Biot-Savart equation together with
a surface integral equation for the electric potential at the vacuum
boundaries. For time-dependent solutions, the model has to be completed
with an integral equation for the magnetic vector potential. All field
quantities are expanded in harmonic modes ($\sim\exp\left({\textrm{i}}\, m\varphi\right)$)
in azimuthal direction and vary in time $t$ according to $\exp\left(pt\right)$
with a constant $p$ that is, in general, complex. Then, there are
two ways to solve the integral equation system. For steady eigenfields
(i.e. marginal eigenfields which are non-oscillatory) it is treated
as an eigenvalue equation for the critical value of $\alpha$. For
unsteady eigenfields (including marginal eigenfields which are oscillatory)
the integral equation system is treated as an eigenvalue problem in
$p$: dynamo solutions corresponding to exponentially growing magnetic
fields are characterised by a positive real part of $p$ . In Appendix
C more details about this numerical approach are given.

We stress that the $\alpha$ effect has been determined under the
assumption of an axisymmetric mean magnetic field with $m=0$. Using
the same $\alpha$ effect for other $m$ modes is an approximation
which is valid only if $m\ll n$, where $n$ is the number of pairs
of rolls. In that case the azimuthal variation of $\bmB$ is weak
compared to the one of $\bu$.

\subsection{Karlsruhe geometry}

It is well known that the main generation mechanism of the Karlsruhe
dynamo experiment is an $\alpha$ effect which maintains, in the marginal
case, a steady equatorial dipole (ED) field, i.e. a mode with $m=1$.
For numerical studies, a simplified geometry has been assumed in form
of a finite cylinder with height $H$ and radius $R$. We use this
simplified geometry and the $\bscE$ given by (\ref{eq:efm-Karlsruhe})
to compute dynamo solutions for different ratios $H/R$. In figure
1 we represent the threshold $C_{\alpha}^{c}$ of the dynamo number
$C_{\alpha}=\mu\sigma R\alpha$ corresponding to $\Re\{ p\}=0$. This
is done for the two leading axisymmetric modes with $m=0$, i.e. for
the axial dipole (AD) and the quadrupole (Q), as well as for the first
non-axisymmetric mode ($m=1$) which represent an equatorial dipole
(ED). All these modes are steady at the marginal point.

\begin{figure}[h]
\begin{center}\includegraphics[%
  width=7cm,
  angle=-90]{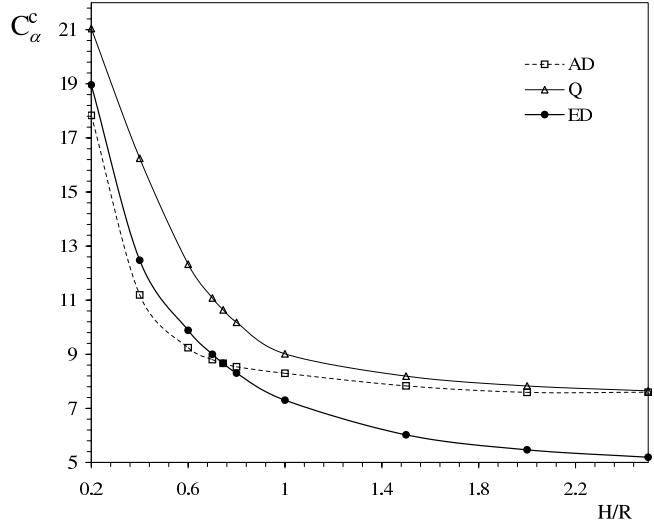}\end{center}

\caption{Critical dynamo number $C_{\alpha}^{c}$ as a function of the aspect
ratio $H/R$ for a Karlsruhe-type dynamo. The two first axisymmetric
($m=0$) eigenmodes AD (axial dipole) and Q (quadrupole) are compared
with the first non-axisymmetric ($m=1$) eigenmode ED (equatorial
dipole).}
\end{figure}

We have found a critical aspect ratio $H/R=0.75$ which distinguishes
between dominant ED and AD fields. Above this critical value ED fields
are dominant while below this value AD fields are dominant. Actually,
the critical aspect ratio of 0.75 is not very far from the experimental
one, which is 0.83.

\subsection{Ring of rolls}

In the following, we investigate anisotropic $\alpha^{2}$ dynamos
in an annular space defined by a gap width of $2\delta R$ with $\delta<1$.
Note that in the following $R$ refers to the radius in the middle
of the gap, and not to the outer radius. We consider both the $z$-independent
case with $\alpha$ given by (\ref{eq:matrix-alpha}) and the $z$-dependent
case with $\alpha$ given by (\ref{eq:matrix-alpha-z}). In each case
we have considered two types of flow distinguished by the radial dependence
of their vertical velocities as defined in Appendix A.2. We have called
them FW1 and FW2. We computed the critical value $C_{\alpha}^{c}$
of the dynamo number $C_{\alpha}=\mu\sigma R\tilde{\alpha}$ where
$\tilde{\alpha}$ stands for $\sqrt{\left\langle \alpha_{\varphi\varphi}^{2}\right\rangle }$
with $\left\langle \cdot\cdot\cdot\right\rangle $ understood as an
average over $s$. According to Avalos \textit{et al.} (2007) the
relation between $\tilde{\alpha}$ and the real velocity of the flow
is given, under the first order smoothing approximation, by $\tilde{\alpha}\approx(\eta\delta/R)R_{m\perp}R_{m\parallel}$.
The quantities $R_{m\perp}=u_{0\perp}R/\eta$ and $R_{m\parallel}=u_{0\parallel}R/\eta$
are the magnetic Reynolds numbers expressed in terms of the characteristic
velocities in the horizontal (i.e. perpendicular to the $z$-axis)
and in the axial (i.e. parallel to the $z$-axis) direction, respectively.

\subsubsection{$z$- independent case}

In figure 2 the dynamo threshold is plotted in dependence on $H/R$
for both flows FW1 and FW2 for $\delta=0.5$ and $n=4$. Quite similar
to the Karlsruhe dynamo case, a critical aspect ratio $H/R$ is also
found here for both flows, which distinguishes between dominant ED
and AD fields. Another critical value of $H/R$ is found where the
second (i.e. subdominant) eigenmode is switching between AD and Q.

For a given ratio $H/R$ we found that the dynamo threshold increases
monotonically when we reduced the magnitudes of $\alpha_{ss}$ and
$\alpha_{\varphi\varphi}$ while keeping $\alpha_{z\varphi}$ unchanged.
This is related to the impossibility of having dynamo action with
a z-independent horizontal flow only.

\begin{center}%
\begin{figure}[h]
\begin{center}\subfigure[]{\includegraphics[%
  clip,
  width=7cm,
  keepaspectratio]{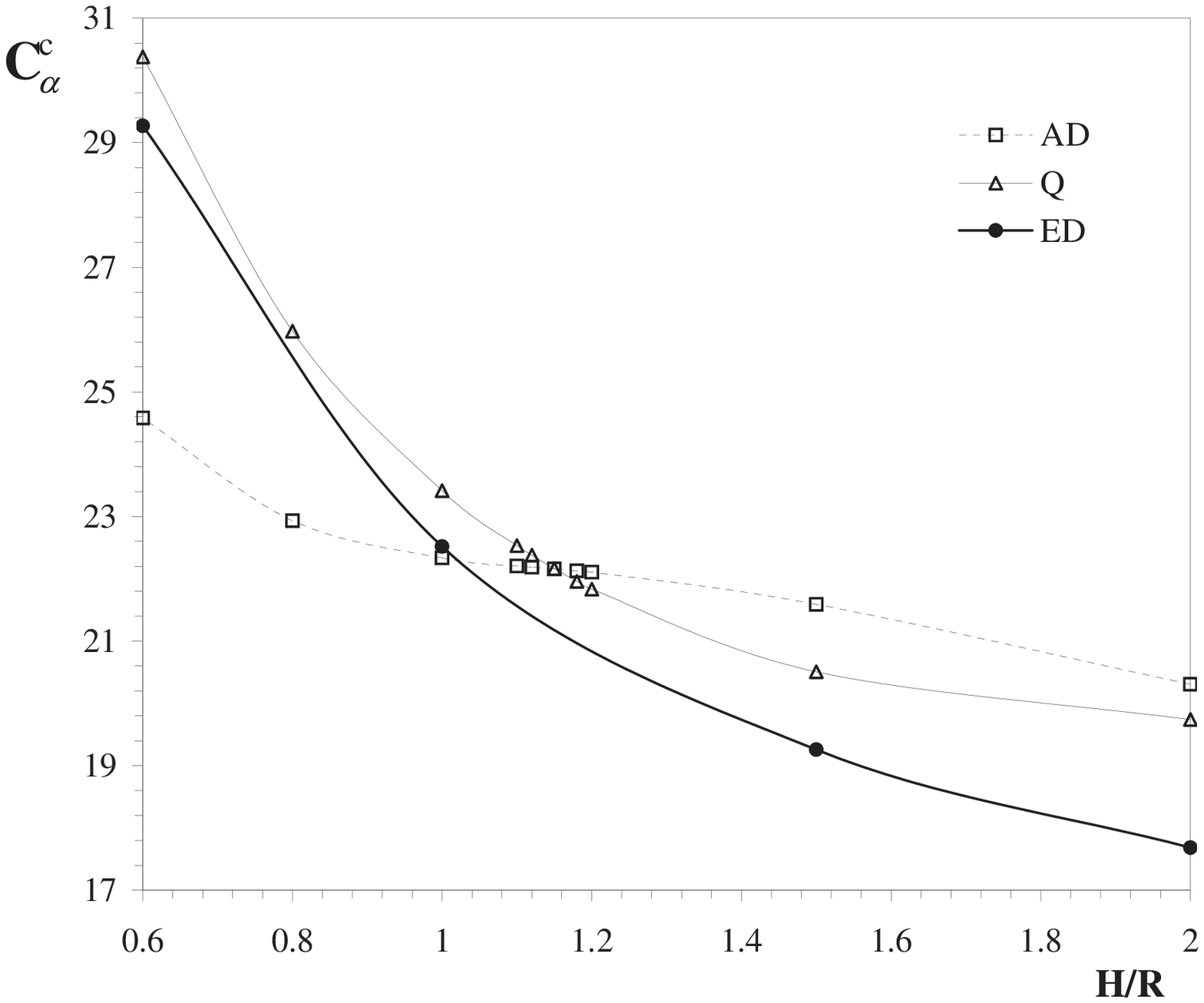}} $\qquad$\subfigure[]{\includegraphics[%
  clip,
  width=7cm,
  keepaspectratio]{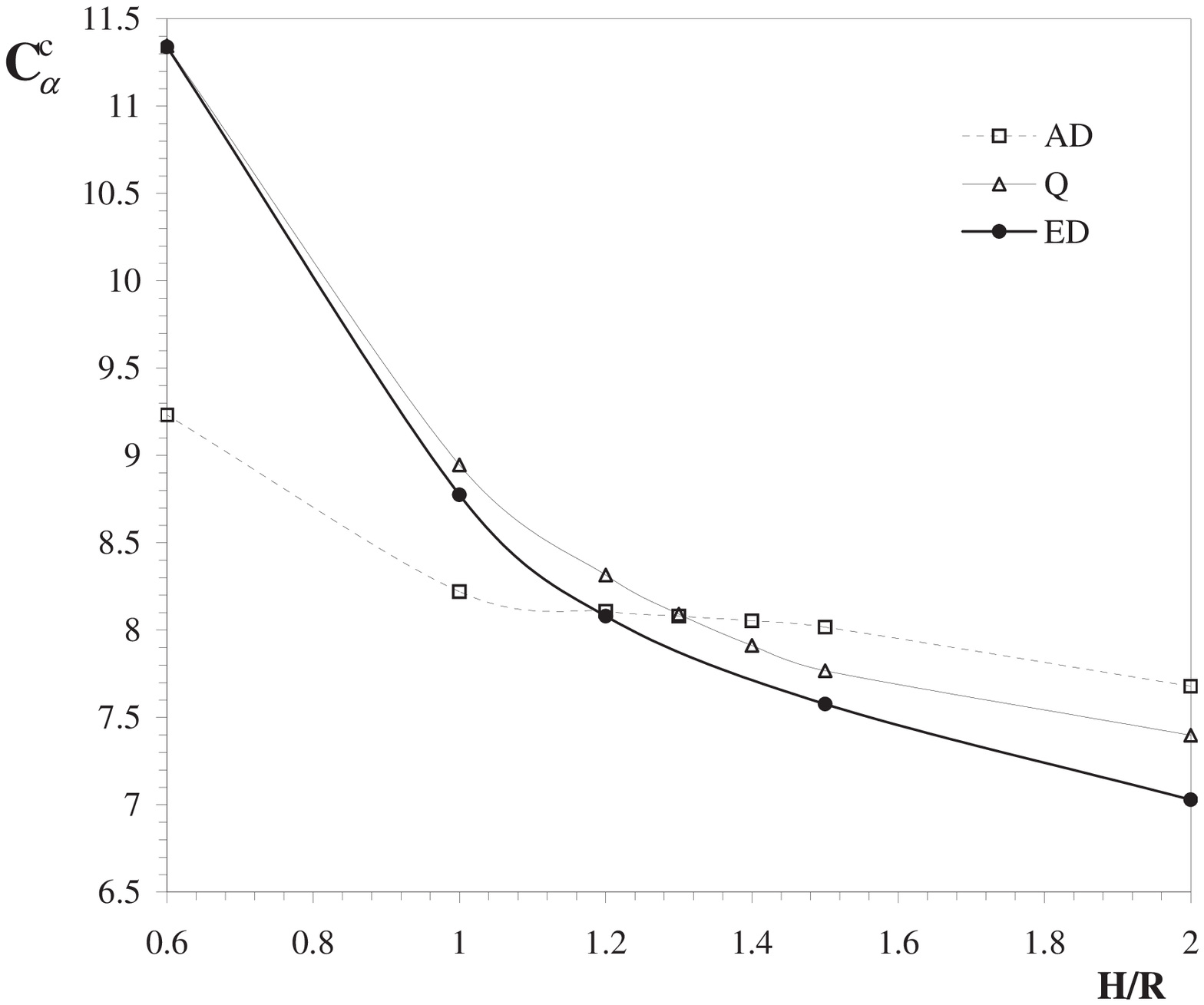}}\end{center}

\caption{Critical dynamo number $C_{\alpha}^{c}$ as a function of the aspect
ratio $H/R$ for an ${\alpha}$ matrix according to (5) with $\delta=0.5$.
The two first axisymmetric ($m=0$) eigenmodes AD (axial dipole) and
Q (quadrupole) are compared with the first non-axisymmetric ($m=1$)
eigenmode ED (equatorial dipole). Plot (a) for FW1 and (b) for FW2.}
\end{figure}
\end{center}

In figure 3, the rescaled dynamo threshold $\delta C_{\alpha}^{c}$
is plotted in dependence on $\delta$ for both flows FW1 and FW2.
We introduce here another distinction between the case of ''free
rolls'' (for which the number of pairs of rolls is kept equal to
4 independently of the value of $\delta$) and the case of ''compact
rolls'' (for which the rolls have the same extension in azimuthal
and radial direction and the number of pairs of rolls scales like
$n=\pi/2\delta$). In neither case was there any indication for a
critical value of $\delta$ below which the dominant $m=1$ mode is
clearly replaced by a dominant $m=0$ mode. However, for small values
of $\delta$, the values of $\delta C_{\alpha}^{c}$ for the $m=0$
mode come very close to those of the $m=1$ mode.

\begin{figure}[h]
\begin{center}\subfigure[]{\includegraphics[%
  clip,
  width=7cm,
  keepaspectratio]{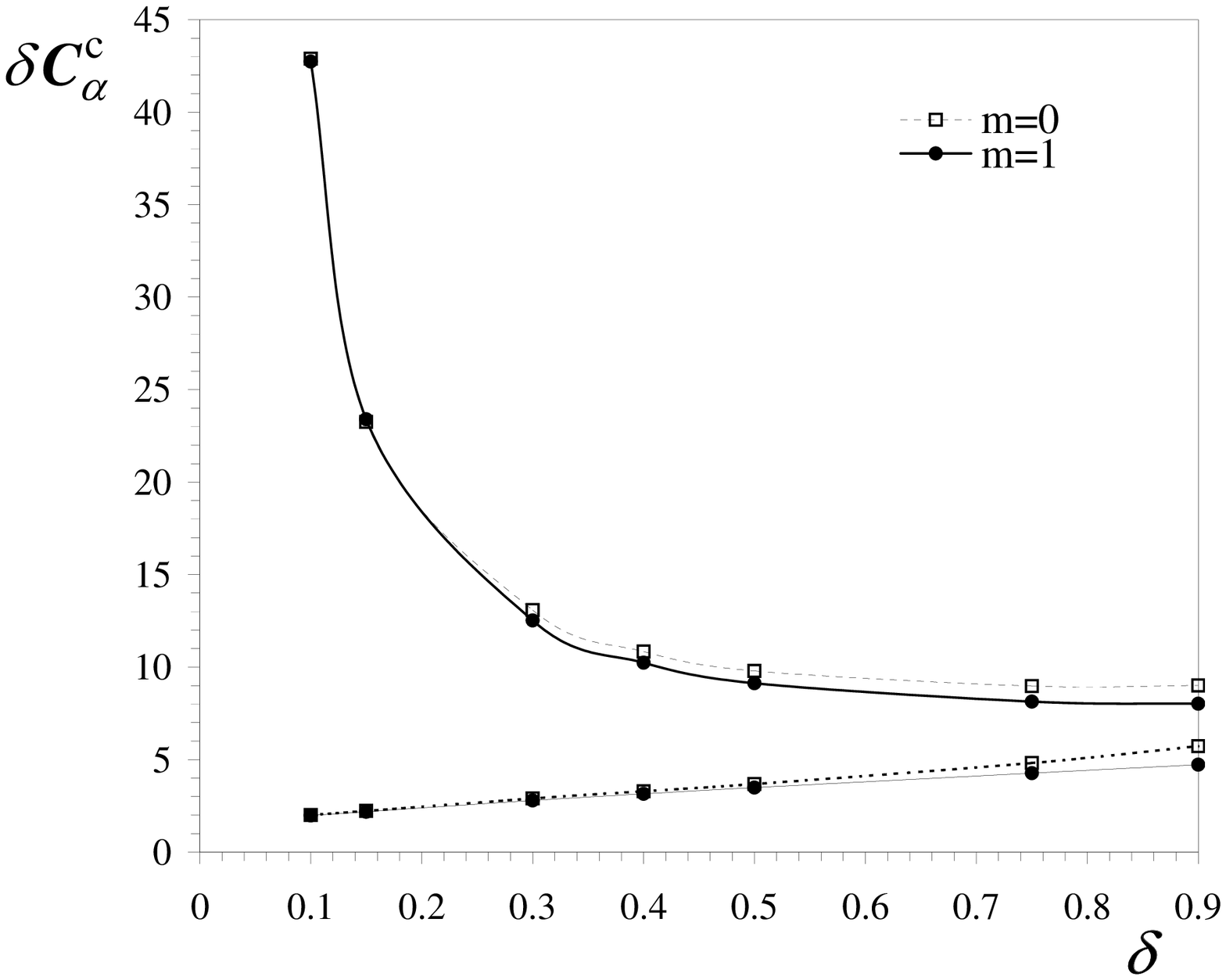}} $\qquad$\subfigure[]{\includegraphics[%
  clip,
  width=7cm,
  keepaspectratio]{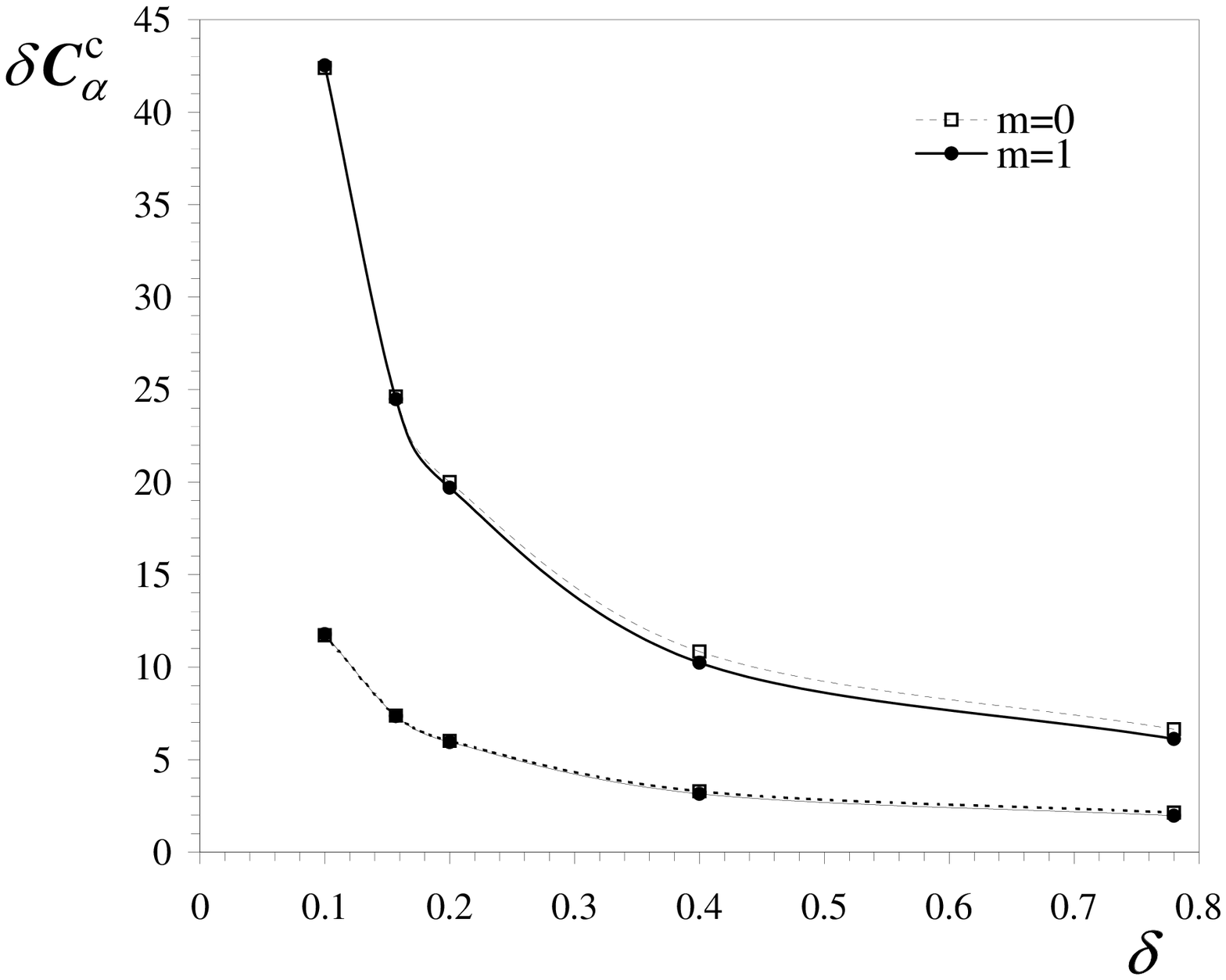}}\end{center}

\caption{The rescaled dynamo threshold $\delta\, C_{\alpha}^{c}$ in dependence
on $\delta$ for FW1(upper curves) and FW2(lower curves), and fix
aspect ratio $H/R=2$. The dashed (solid) line corresponds to axisymmetric
(non axisymmetric) fields. Plot (a) for the case of ''free rolls''
and (b) for the case of ''compact rolls''. }
\end{figure}

In the case of free rolls it is remarkable that $\delta C_{\alpha}^{c}$
decreases with $\delta$ for FW1 and increases for FW2. The geometries
of the magnetic field produced by FW1 and FW2 for $\delta=0.3$ have
indeed different symmetries. This is illustrated in figures 4 and
5 where poloidal vectors and azimuthal contour of the magnetic field
are plotted. On the other hand the symmetries are similar for $\delta=0.9$.

\begin{figure}[h]
\begin{center}\begin{tabular}{>{\centering}m{1cm}>{\centering}m{5cm}m{0.5cm}>{\centering}m{5cm}}
$\delta$&
$\qquad\qquad$$m=0$ &
&
$\qquad\qquad$$m=1$ \tabularnewline
0.9&
\includegraphics[%
  clip,
  width=6cm]{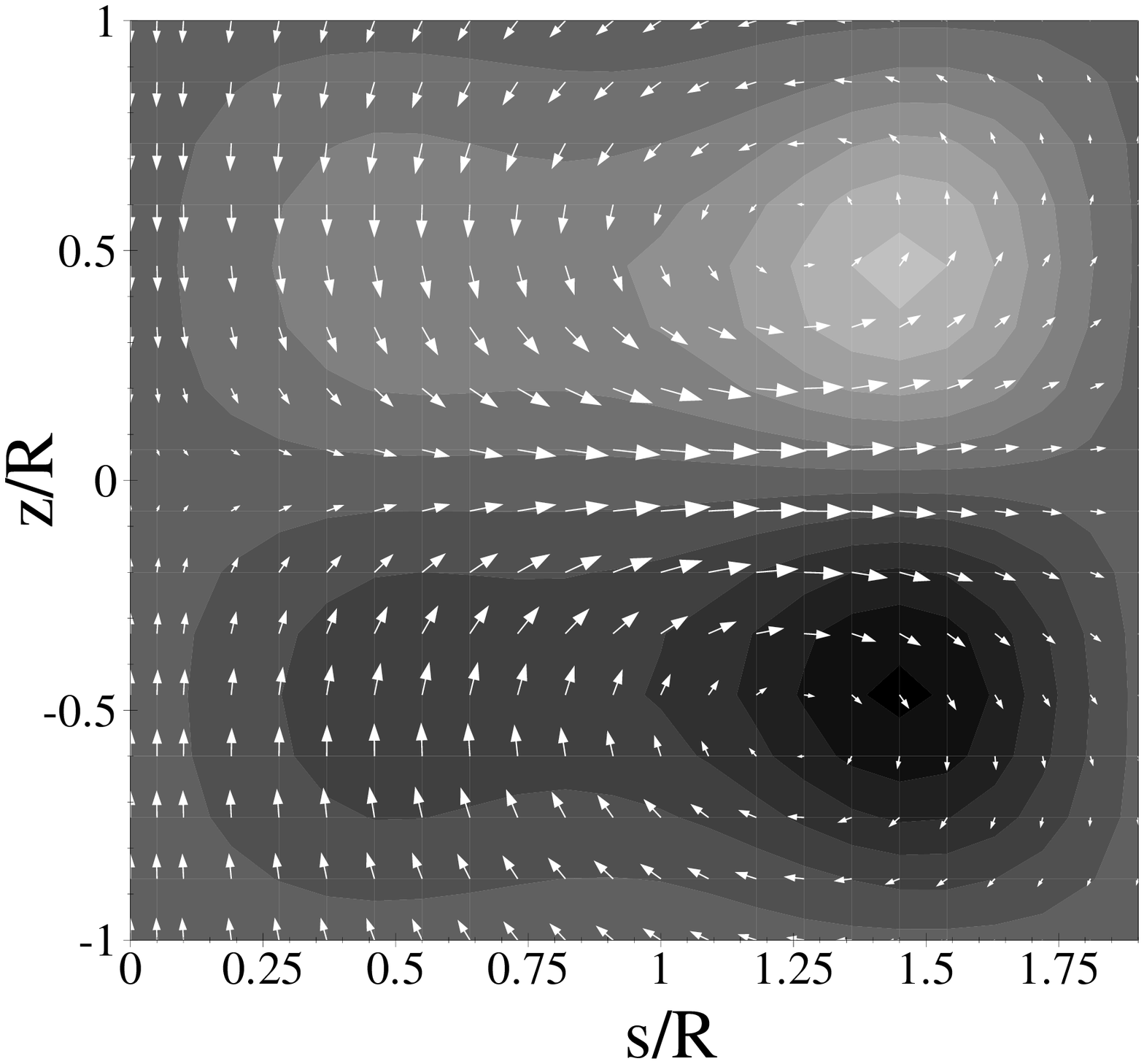}&
&
\includegraphics[%
  clip,
  width=6cm]{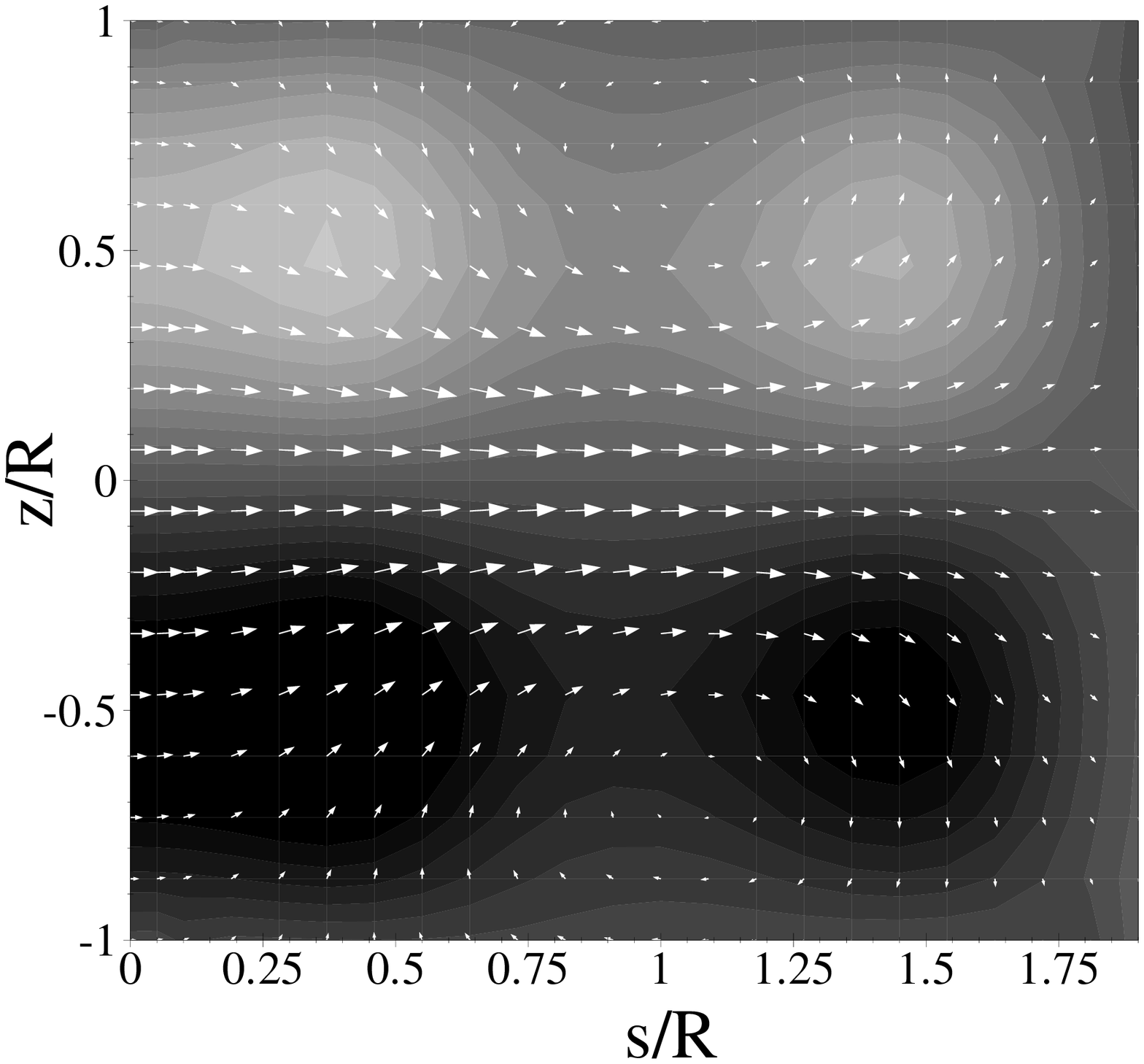}\tabularnewline
0.3&
\includegraphics[%
  clip,
  width=6cm]{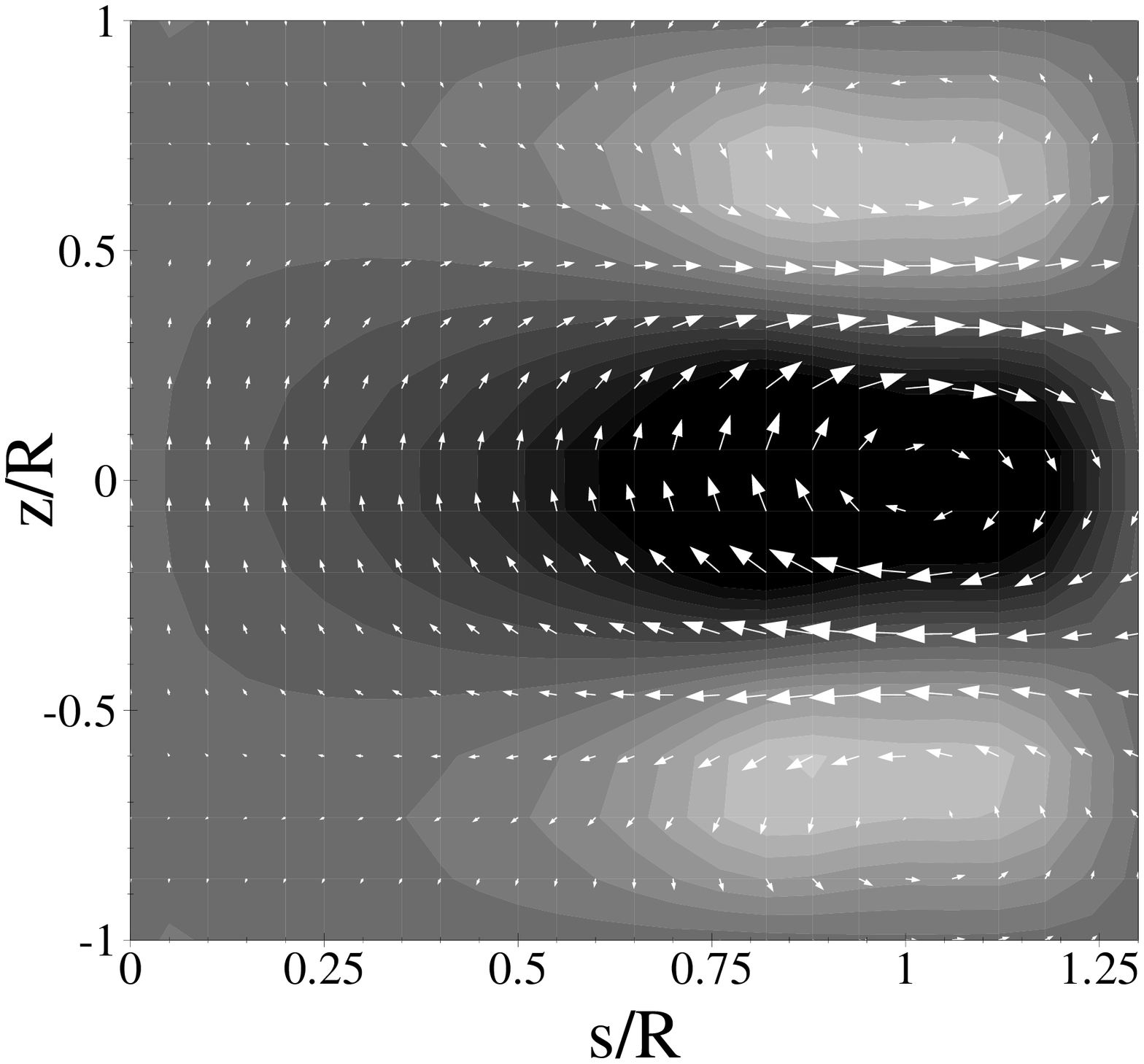}&
&
\includegraphics[%
  clip,
  width=6cm]{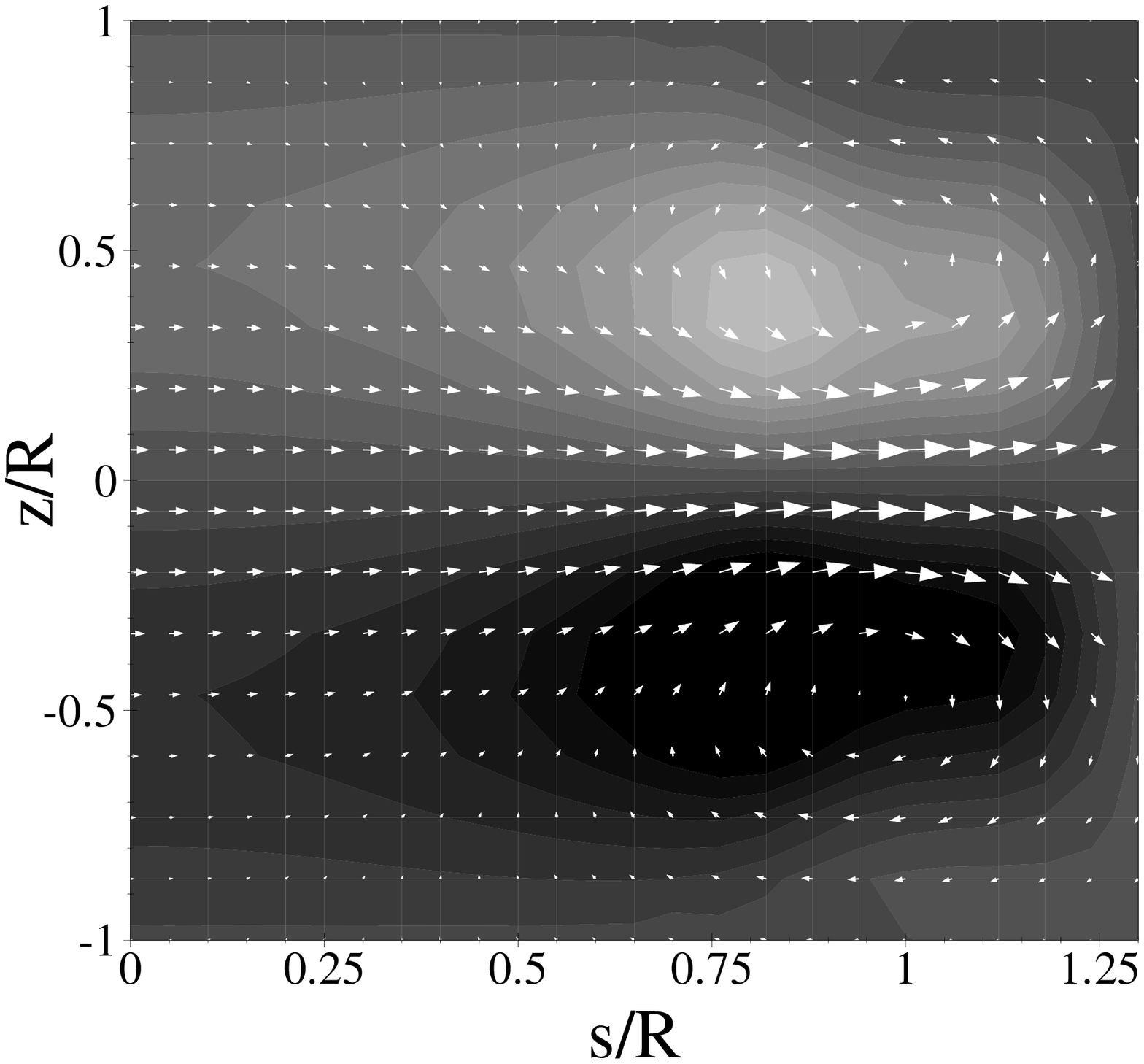}\tabularnewline
\end{tabular}\end{center}

\caption{Poloidal field component together with contour plots of azimuthal
field for FW1 with free rolls, for $\delta=0.3$, $\delta=0.9$, $m=0$,
$m=1$. Ordinate axis runs from $-H/2R$ to $H/2R$, corresponding
to $H/R=2$. For the non-axisymmetric case ($m=1$) the plot represents
the numerical solution in the meridional plane at $\varphi=0$. However,
one should notice the degeneration of the eigenvalue problem with
respect to any rotation of the eigenmode in $\varphi$-direction. }
\end{figure}
\begin{figure}[h]
\begin{center}\begin{tabular}{>{\centering}m{1cm}>{\centering}m{5cm}m{0.5cm}>{\centering}m{5cm}}
$\delta$&
$\qquad\qquad$$m=0$ &
&
$\qquad\qquad$$m=1$ \tabularnewline
0.9&
\includegraphics[%
  clip,
  width=6cm]{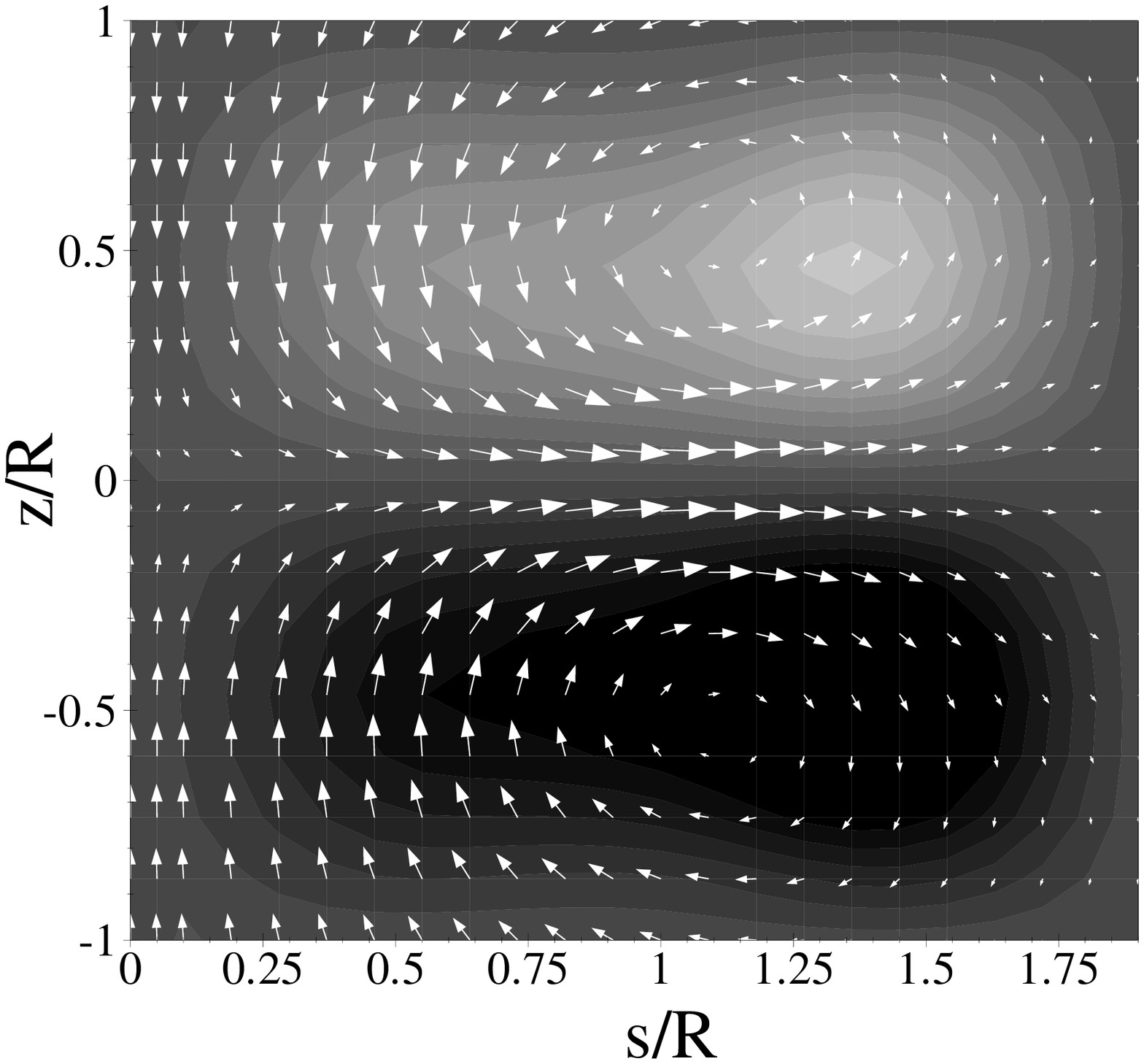}&
&
\includegraphics[%
  clip,
  width=6cm]{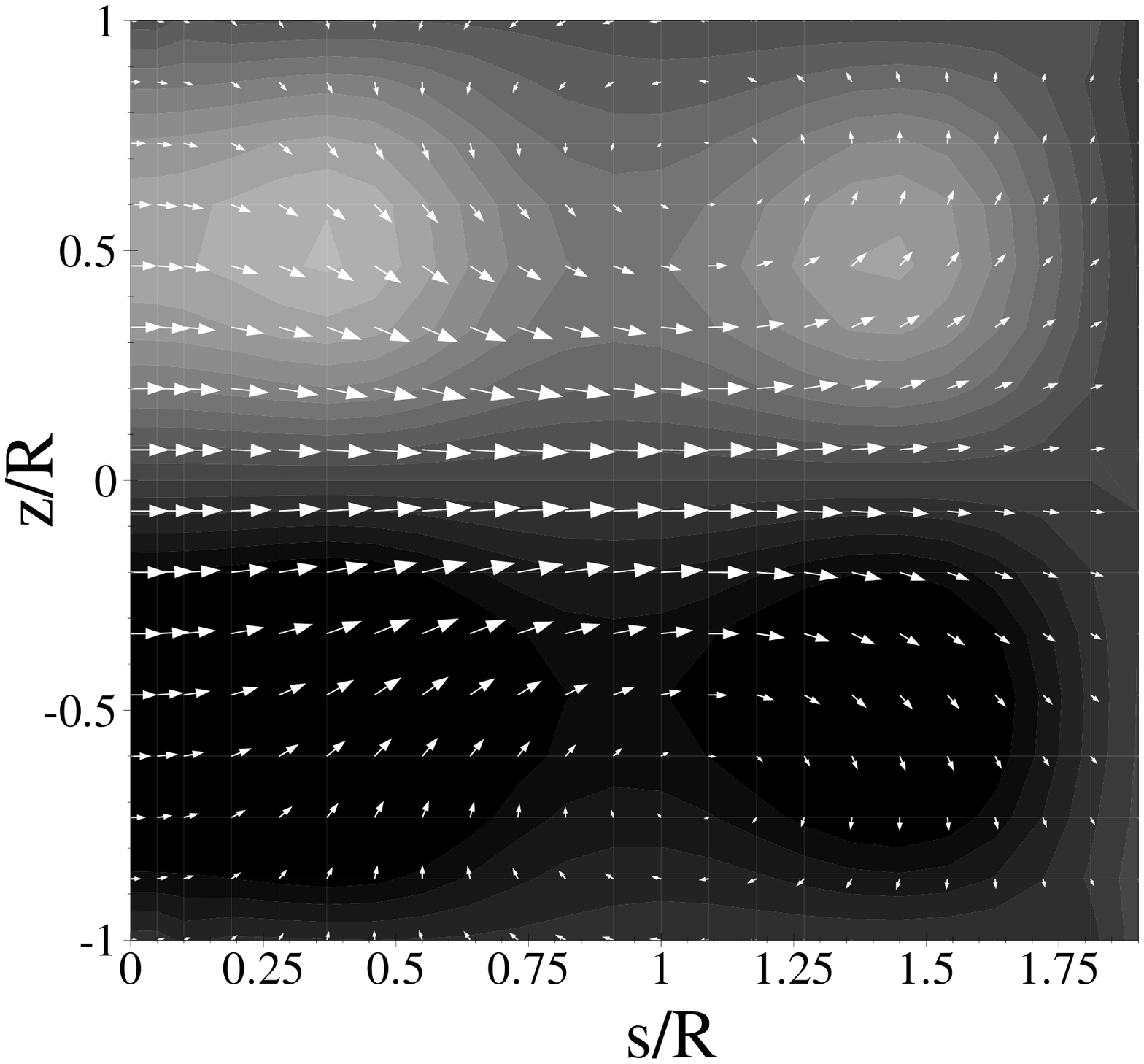}\tabularnewline
0.3&
\includegraphics[%
  clip,
  width=6cm]{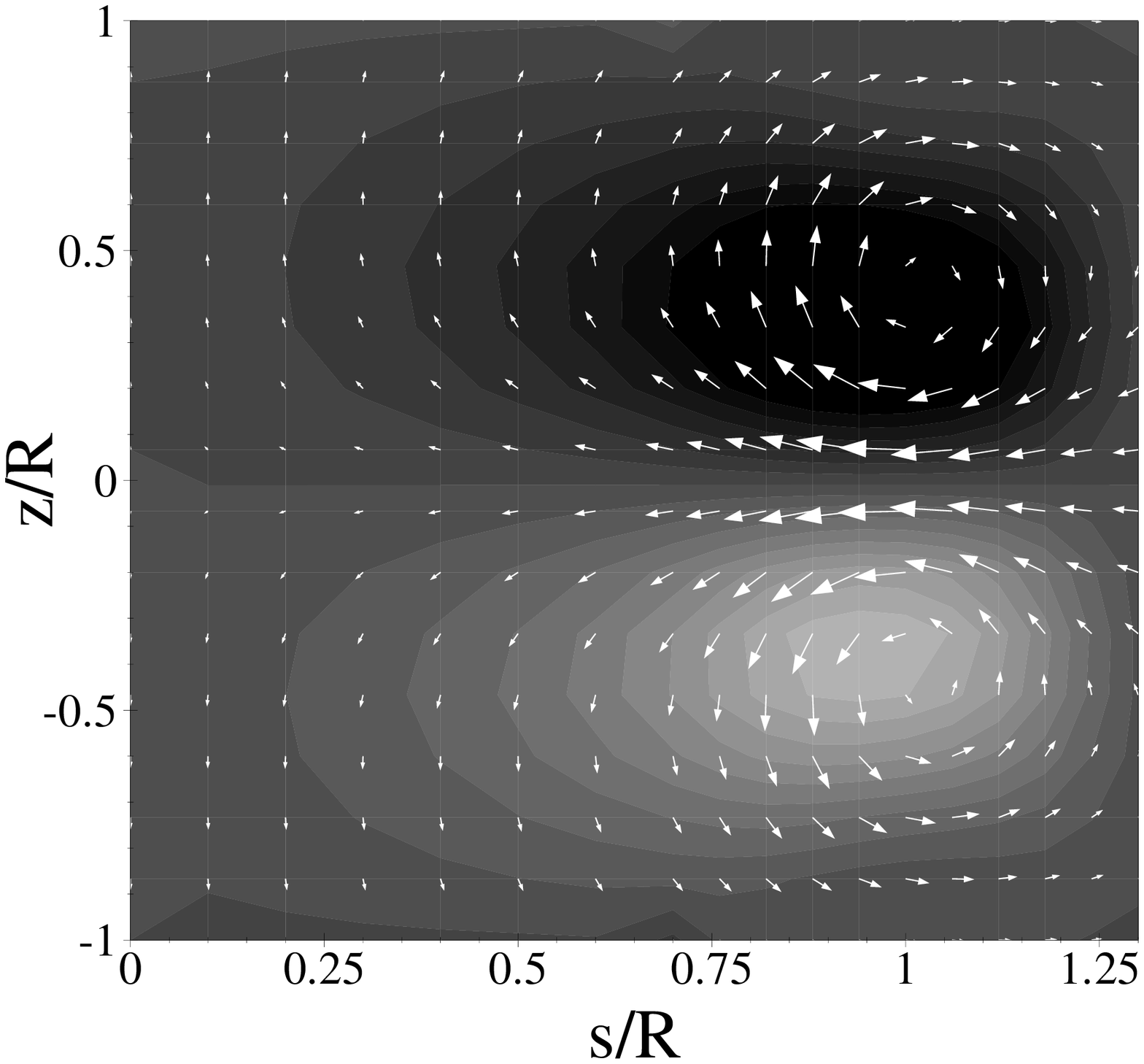}&
&
\includegraphics[%
  clip,
  width=6cm]{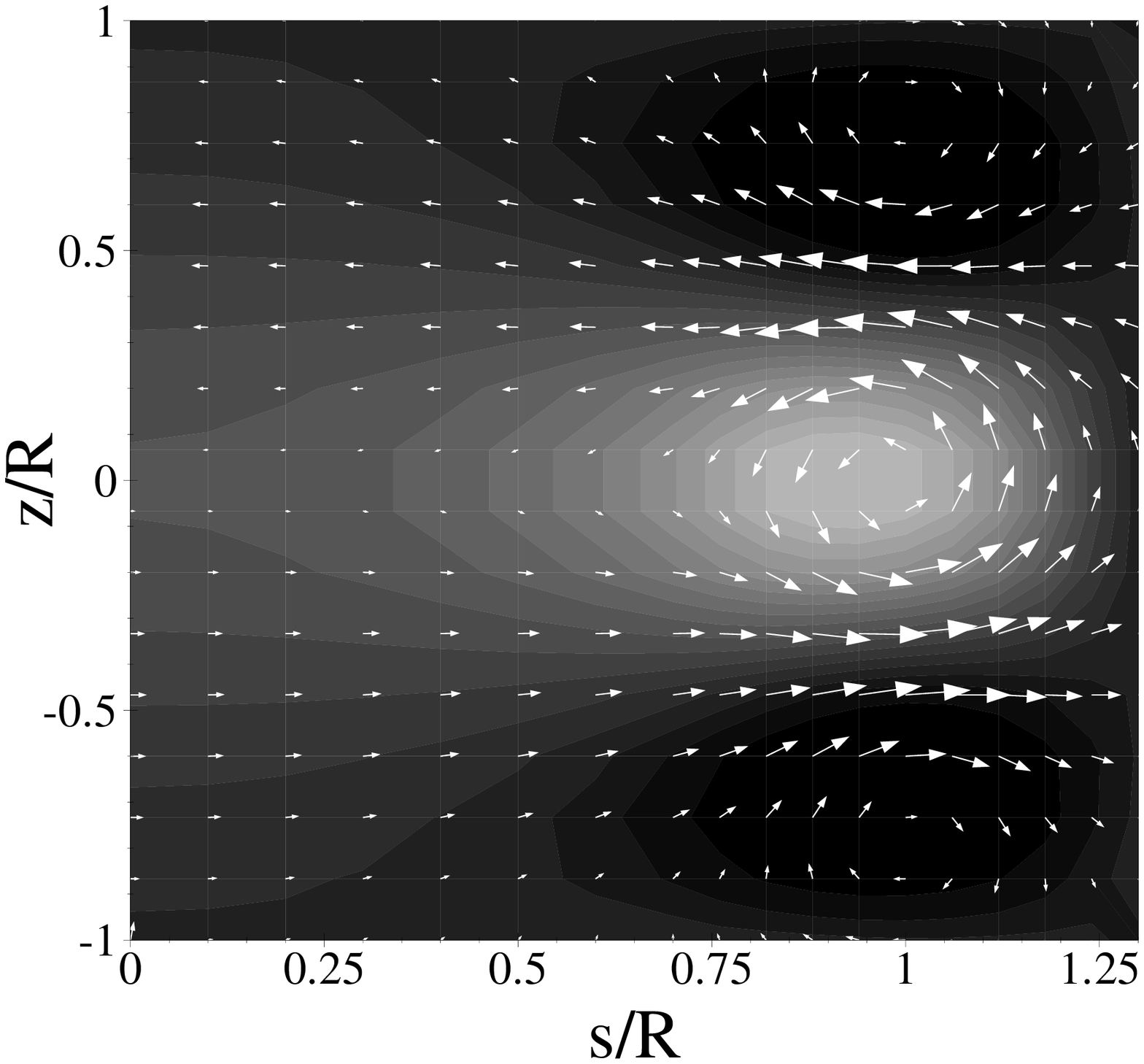}\tabularnewline
\end{tabular}\end{center}

\caption{Same caption as figure 4 but for FW2. }
\end{figure}

\subsubsection{$z$-dependent case}

In figure 6, $C_{\alpha}^{c}$ is plotted in dependence on $H/R$
in the $z$ dependent case for $\delta=0.5$ and $n=4$. We find now
that oscillatory non-axisymmetric fields are always the most easily
excitable solutions for both flow types FW1 and FW2. However, the
axisymmetric solutions are getting closer to non-axisymmetric ones
as $H/R$ is reduced. For the FW1 flow a transition between steady
and oscillatory magnetic fields is observed for the mode $m=0$ at
a certain value $H/R\sim1$ (the precise value could not be determined
since the numerical solution of the problem is quite time consuming)
. In all cases non-dipolar fields only were found. %
\begin{figure}[h]
\begin{center}\subfigure[]{\includegraphics[%
  width=7cm,
  keepaspectratio]{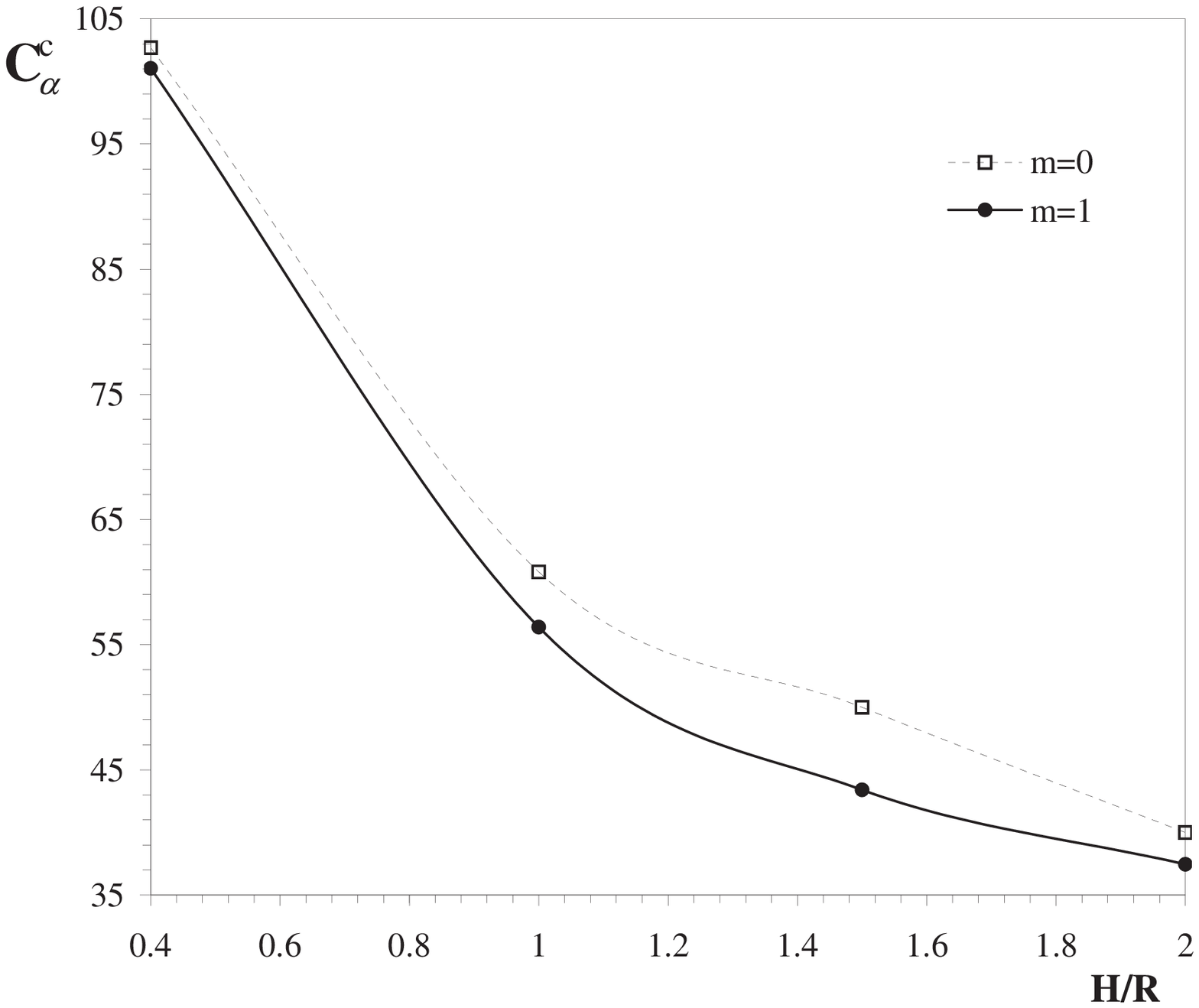}} $\quad$\subfigure[]{\includegraphics[%
  clip,
  width=7cm,
  keepaspectratio]{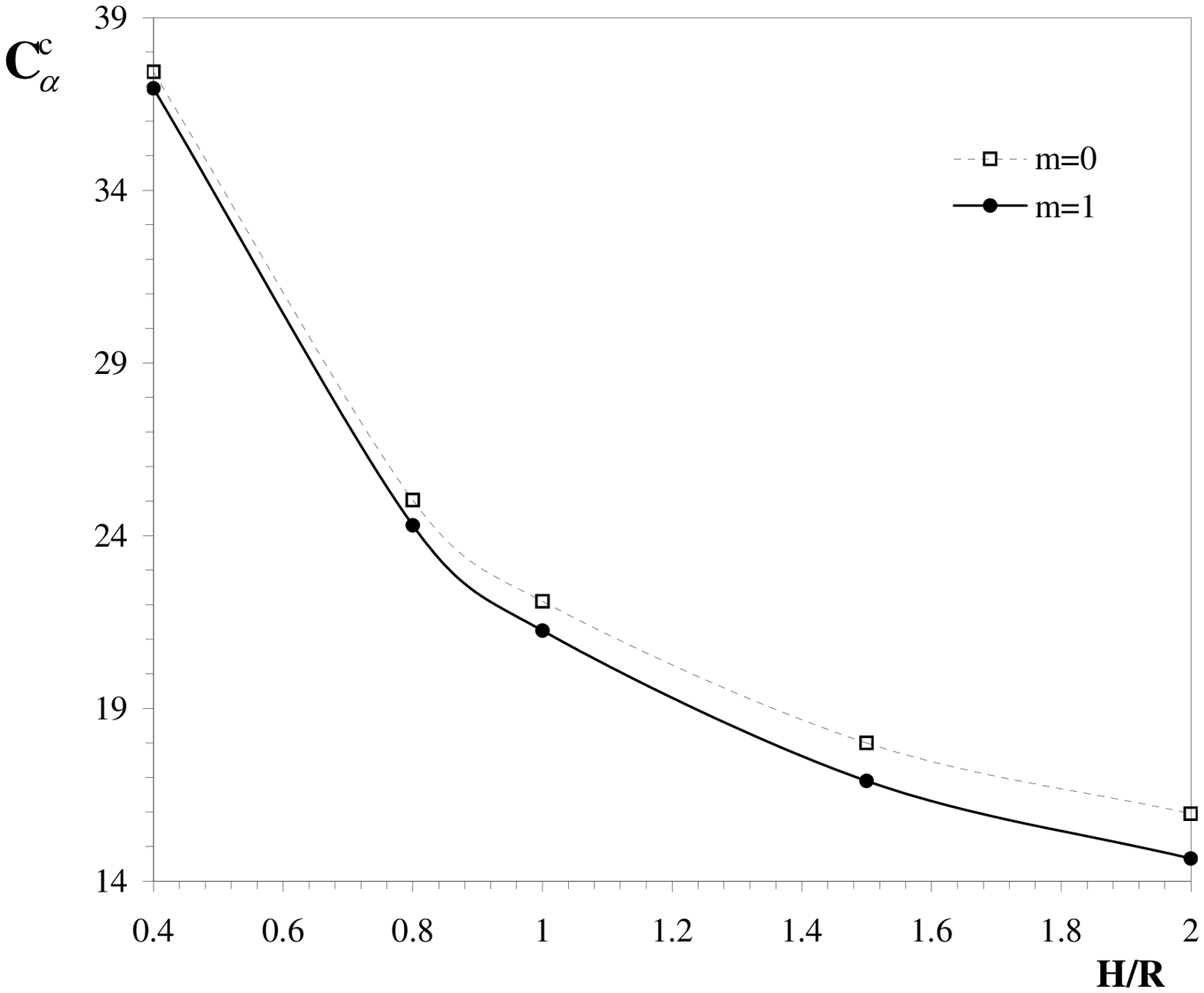}}\end{center}

\caption{Critical dynamo number $C_{\alpha}^{c}$ as a function of cylinder
aspect ratio $H/R$ for an $\alpha$ matrix according to (6). (a)
FW1 and (b) FW2. The dashed (solid) line corresponds to axisymmetric
(non axisymmetric) fields. Solutions in (a) are always oscillatory
except for $m=0$ and higher values of $H=1.04$ where solutions are
steady. Solutions in (b) are always oscillatory.}
\end{figure}

\section{Conclusions}

We have explored the influence of geometrical parameters on spatial
structure and temporal variations of magnetic fields generated by
kinematic anisotropic $\alpha^{2}$ dynamos working in a finite cylinder.
The $\alpha$ coefficients were calculated for specific flow patterns,
following the lines of mean field concept, and the corresponding dynamo
solutions were calculated using the integral equation approach. The
obtained results show that this kind of dynamos can switch from dominant
equatorial dipoles to dominant axial dipoles just by reducing the
aspect ratio of the cylinder. This transition occurs for quite different
forms of $\alpha$: constant, as in the Karlsruhe dynamo experiment,
or having a purely radial dependence, as the one obtained in a flow
described by axially invariant helical columns. On the other hand,
such a transition does not occur when the relative gap width $\delta$
is reduced (at least not for the considered aspect ratio). When $\alpha$
has an additional axial dependence, dominant dynamo solutions are
only oscillatory $m=1$ modes. In addition for the $m=0$ mode both
steady and oscillatory solutions were obtained.

\section*{Acknowledgments}

This work was supported by Deutsche Forschungsgemeinschaft in frame
of SFB 609 and Grant No. GE 682/14-1. We are grateful to Karl-Heinz
R\"{a}dler for many valuabel comments on the paper.

\section*{Appendix A: Specification of the velocity field}

\subsection*{A.1 General assumptions}

We specify the motion of an incompressible conducting fluid $\bu$,
so that it corresponds to a ring of columnar vortices.The ring is
coated by an interval $1-\delta\leq s/R\leq1+\delta$ with $\delta<1$.
Outside this interval the fluid is assumed to be at rest. It is assumed
that $\bu$ is steady, $z$-independent and varies with $\varphi$
like $\exp(\iu n\varphi)$, where $n$ is the number of vortex pairs.
We use the representation 

\begin{eqnarray*}
\bu & = & -\bnab\times(\mathrm{\mathbf{e}}_{z}\times\bnab\Phi)-\mathrm{\mathbf{e}}_{z}\times\bnab\Psi,\\
\qquad\qquad\qquad\qquad\Phi & = & u_{0\parallel}R^{2}\phi(s)\cos(n\varphi)\,,\quad\Psi=u_{0\perp}R\,\psi(s)\cos(n\varphi).\qquad\;\;{\rm (A.1)}\end{eqnarray*}
 The two terms on the right-hand side of $\bu$ correspond to the
vertical (poloidal) and horizontal (toroidal) parts of the velocity.
The constant quantities $u_{0\perp}$ and $u_{0\parallel}$ define
the intensity of the considered flow. We further express $\bu$ by
\[
\qquad\qquad u_{s}=\hat{u}_{s}(s)\sin(n\varphi),\quad u_{\varphi}=\hat{u}_{\varphi}(s)\cos(n\varphi),\quad u_{z}=\hat{u}_{z}(s)\cos(n\varphi).\qquad\;{\rm (A.2)}\]
The connection between (A.1) and (A.2) is given by \[
\qquad\qquad\hat{u}_{s}=-u_{0\perp}R\,\frac{n}{s}\psi\,,\quad\hat{u}_{\varphi}=-u_{0\perp}R\,\frac{\partial\psi}{\partial s}\,,\quad\hat{u}_{z}=-u_{0\parallel}R^{2}D_{n}\phi\,,\qquad\qquad\,\mathrm{(A.3)}\]
where $D_{n}\phi=s^{-1}\partial_{s}(s\,\partial_{s}\,\phi)-(n/s)^{2}\phi.$

\subsection*{A.2 Specific examples}

We consider two flows which differ only in the radial dependence of
$u_{z}$. The first flow (FW1) is defined by \begin{eqnarray*}
 &  & \psi=C_{\psi}\left(1-\xi^{2}\right)^{3},\quad\hat{u}_{z}/u_{0\parallel}=C_{z}\left(1-\xi^{2}\right)^{2},\quad\xi=\frac{\left(s/R\right)-1}{\delta}\,,\quad\mbox{if}\quad|\xi|<1\\
\qquad &  & \phi=\psi=0\quad\mbox{otherwise}\,.\qquad\qquad\qquad\qquad\qquad\qquad\qquad\qquad\qquad\qquad\:\mathrm{(A.4)}\end{eqnarray*}
 The second one (FW2) by \begin{eqnarray*}
 &  & \psi=C_{\psi}\left(1-\xi^{2}\right)^{3},\quad\phi=C_{\phi}\left(1-\xi^{2}\right)^{3},\quad\xi=\frac{\left(s/R\right)-1}{\delta}\,,\quad\mbox{if}\quad|\xi|<1\\
\qquad &  & \phi=\psi=0\quad\mbox{otherwise}\,.\qquad\qquad\qquad\qquad\qquad\qquad\qquad\qquad\qquad\qquad\:\mathrm{(A.5)}\end{eqnarray*}
 The factors $C_{\psi}$ , $C_{\phi}$ and $C_{z}$ were chosen such
that the average of $u_{z}/u_{0\parallel}$ over a surface given by
$1-\delta\leq s/R\leq1+\delta$, $-\pi/2n\leq\varphi\leq\pi/2n$ and
$z/R=\mbox{constant}$ as well as the average of $u_{\varphi}/u_{0\perp}$
at $\varphi=0$ over $1\leq s/R\leq1+\delta$ are equal to unity,

\[
C_{\psi}=\delta,\quad C_{z}=\frac{{15\pi}}{16},\]

\begin{flushright}\[
\quad\quad C_{\phi}=\left.{\frac{15\pi\delta^{7}}{n^{2}}}\right/\left[2\delta\left(15-40\delta^{2}+33\delta^{4}\right)+15\left(1-\delta^{2}\right)^{3}\:\log\left(\frac{1-\delta}{1+\delta}\right)\right].\qquad\;\mathrm{(A.6)}\]
\end{flushright}

The flow definitions (A.4) and (A.5) ensure that $u_{s}$, $u_{\varphi}$
and $u_{z}$ are continuous and have continuous derivatives everywhere.
In figure 7 we give an example of both flow geometries.

\begin{center}%
\begin{figure}[h]
\begin{center}\subfigure[]{\includegraphics[%
  width=7cm,
  keepaspectratio]{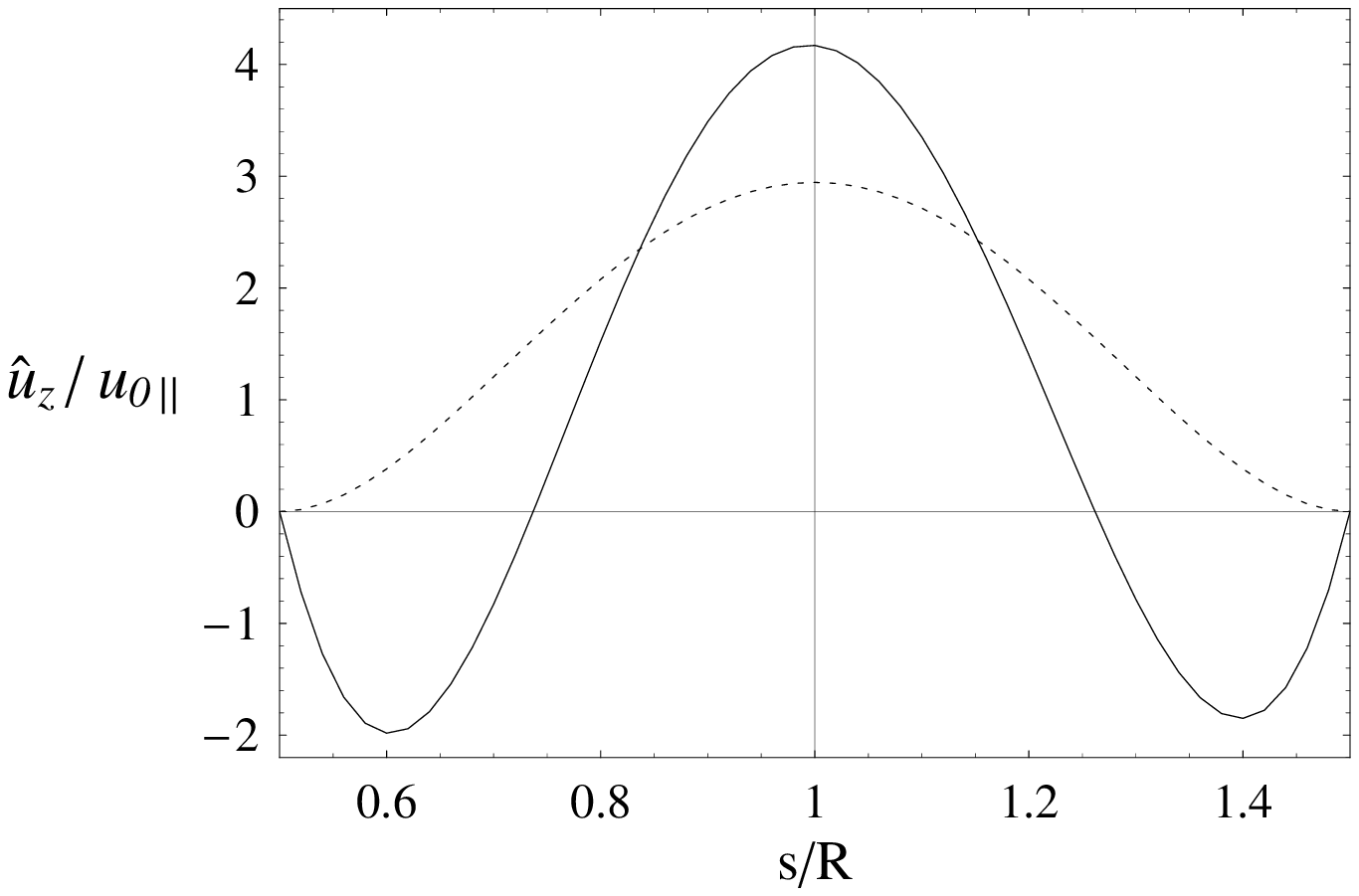}} $\qquad$\subfigure[]{\includegraphics[%
  clip,
  width=4.8cm,
  keepaspectratio]{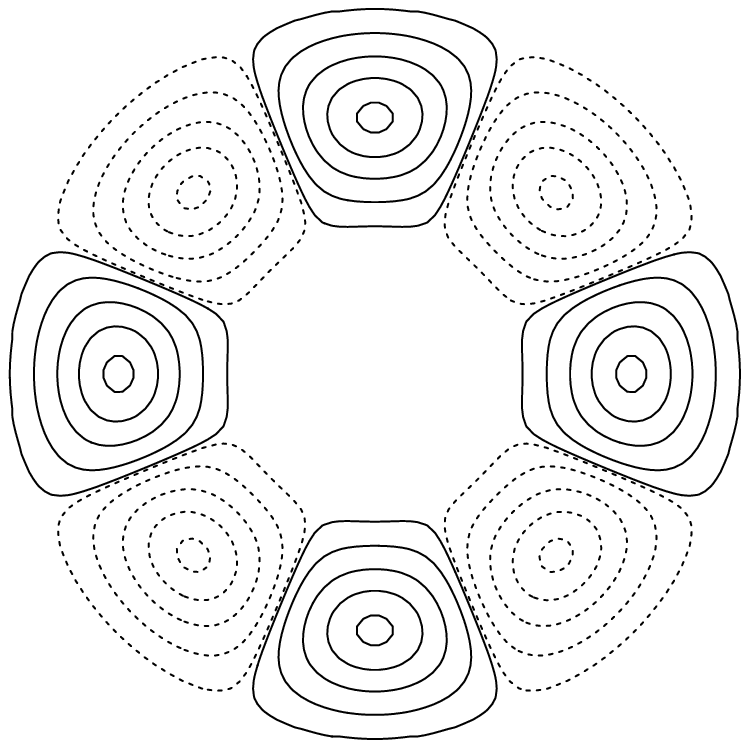}}\end{center}

\caption{Representation of a ring of rolls with four vortex pairs ($n=4$)
and dimensionless annular width $\delta=0.5$. Plot (a) shows radial
profile of the dimensionless vertical velocity in a vortex for first
flow (FW1, dashed line) and second flow (FW2, solid line), while plot
(b) shows streamlines of horizontal fluid motion for both flows. Solid
and dashed lines correspond to opposite circulations.}
\end{figure}
\end{center}

\section*{Appendix B: Determination of $\bscE$}

We consider an electromotive force $\bscE$ generated by a flow structured
in helical columns between two concentric cylinders that was coined
a ''ring of rolls''. Assuming that $\bu$ and $\bB$ do not depend
on $z$, we look for representations of $\bscE$ in the general form
\[
\qquad\qquad\qquad\qquad\quad{\mathcal{{E}}}_{\kappa}(s)=\int_{0}^{\infty}K_{\kappa\lambda}(s,s')\,\mB_{\lambda}(s')\, s'\,\dd s'\,,\qquad\qquad\qquad\qquad\quad\:{\rm (B.1)}\]
 where $\kappa$ and $\lambda$ stand for $s$, $\varphi$ or $z$.
Using a Taylor expansion of $\mB_{\lambda}$ , we write the last equation
as \[
\qquad\qquad\qquad\qquad\quad{\mathcal{{E}}}_{\kappa}(s)=\alpha_{\kappa\lambda}(s)\,\mB_{\lambda}(s)+\beta_{\kappa\lambda s}(s)\,\frac{1}{R}\frac{\partial\mB_{\lambda}}{\partial s}(s)+\cdots\qquad\qquad\;\;\quad\mathrm{(B.2)}\]

with \begin{eqnarray*}
\alpha_{\kappa\lambda}(s) & = & \int_{0}^{\infty}K_{\kappa\lambda}(s,s')\, s'\,\dd s',\qquad\qquad\qquad\qquad\qquad\;\:\,\mathrm{(B.3)}\\
\qquad\qquad\qquad\qquad\quad\beta_{\kappa\lambda s}(s) & = & R\int_{0}^{\infty}K_{\kappa\lambda}(s,s')\,(s'-s)\, s'\,\dd s'.\qquad\qquad\qquad\;\;\mathrm{(B.4)}\end{eqnarray*}

The first term on the r.h.s of (B.2) represents the $\alpha$ effect,
the second term represents the $\beta$ effect, which will be omitted
throughout the paper. The kernel $K_{\kappa\lambda}(s,s')$ depends
only on $\bu$. Under the first order smoothing approximation (FOSA),
and using a definition of mean-fields by $\varphi$ averaging, an
analytical expression of $K_{\kappa\lambda}(s,s')$ was found in Avalos
\textit{et al.} (2007). The results are: \begin{eqnarray*}
2K_{ss}(s,s') & = & -\frac{R^{2}}{\eta}\,\left(\frac{\partial h_{n}}{\partial s'}(s,s')\,\hat{u}_{\varphi}(s)\,\hat{u}_{z}(s')+\frac{\partial h_{n}}{\partial s}(s,s')\,\hat{u}_{z}(s)\,\hat{u}_{\varphi}(s')\right),\end{eqnarray*}
\begin{eqnarray*}
2K_{\varphi\varphi}(s,s') & = & \frac{R^{2}}{\eta}\, n\,\left(\frac{h_{n}(s,s')}{s}\,\hat{u}_{z}(s)\,\hat{u}_{s}(s')+\frac{h_{n}(s,s')}{s'}\,\hat{u}_{s}(s)\,\hat{u}_{z}(s')\right),\end{eqnarray*}
\begin{eqnarray*}
2K_{z\varphi}(s,s') & = & -\frac{R^{2}}{\eta}\,\left(\frac{\partial h_{n}}{\partial s}(s,s')\,\hat{u}_{s}(s)\,\hat{u}_{s}(s')+n\frac{h_{n}(s,s')}{s}\,\hat{u}_{\varphi}(s)\,\hat{u}_{s}(s')\right),\end{eqnarray*}

\begin{flushright}\begin{eqnarray*}
\qquad\qquad\qquad\qquad\quad K_{s\varphi} & = & K_{\varphi s}=K_{zs}=K_{zz}=0\,.\qquad\qquad\qquad\qquad\quad\;\;\quad{\rm (B.5)}\end{eqnarray*}
\end{flushright}

The coefficients $K_{sz}$ and $K_{\varphi z}$ are not zero, but
the integrals $\int_{0}^{\infty}K_{sz}(s,s')\, s'\,\dd s'$ and $\int_{0}^{\infty}K_{\varphi z}(s,s')\, s'\,\dd s'$
can be shown to vanish. 

The Green's function $h_{n}$ are defined by 

\begin{eqnarray*}
h_{n}(s,s')=\frac{1}{2n}\left(\frac{s'}{s}\right)^{2} & \mbox{for} & s'\leq s\\
\qquad\qquad\qquad\qquad\quad h_{n}(s,s')=\frac{1}{2n}\left(\frac{s}{s'}\right)^{2} & \mbox{for} & s\leq s'\,.\qquad\qquad\qquad\qquad\qquad\mathrm{(B.6)}\end{eqnarray*}

As in Avalos \textit{et al.} (2007), we can further represent \[
\alpha_{\kappa\lambda}=\frac{\eta}{R}\, R_{m\perp}\left\{ \begin{array}{c}
R_{m\perp}\\
R_{m\parallel}\end{array}\right\} \widetilde{\alpha}_{\kappa\lambda}\quad\textrm{if}\quad(\kappa\lambda)=\left\{ \begin{array}{cc}
(z\varphi)\\
(ss), & (\varphi\varphi)\end{array},\right.\;\widetilde{\alpha}_{\kappa\lambda}=0\quad\textrm{otherwise.}\]
\[
\qquad\qquad\qquad\qquad\qquad\qquad\qquad\qquad\qquad\qquad\qquad\qquad\qquad\qquad\qquad\qquad\quad\,\mathrm{(B.7)}\]
where $\widetilde{\alpha}_{\kappa\lambda}$ is a dimensionless quantity
independent of magnetic Reynold numbers $R_{m\perp}=u_{0\perp}R/\eta$
and $R_{m\parallel}=u_{0\parallel}R/\eta$. In figure 8, the $s/R$
profile of the three non-zero dimensionless $\widetilde{\alpha}_{\kappa\lambda}$
coefficients are represented for both flows FW1 and FW2. %
\begin{figure}[h]
\begin{center}\subfigure[]{\includegraphics[%
  width=7.1cm,
  keepaspectratio]{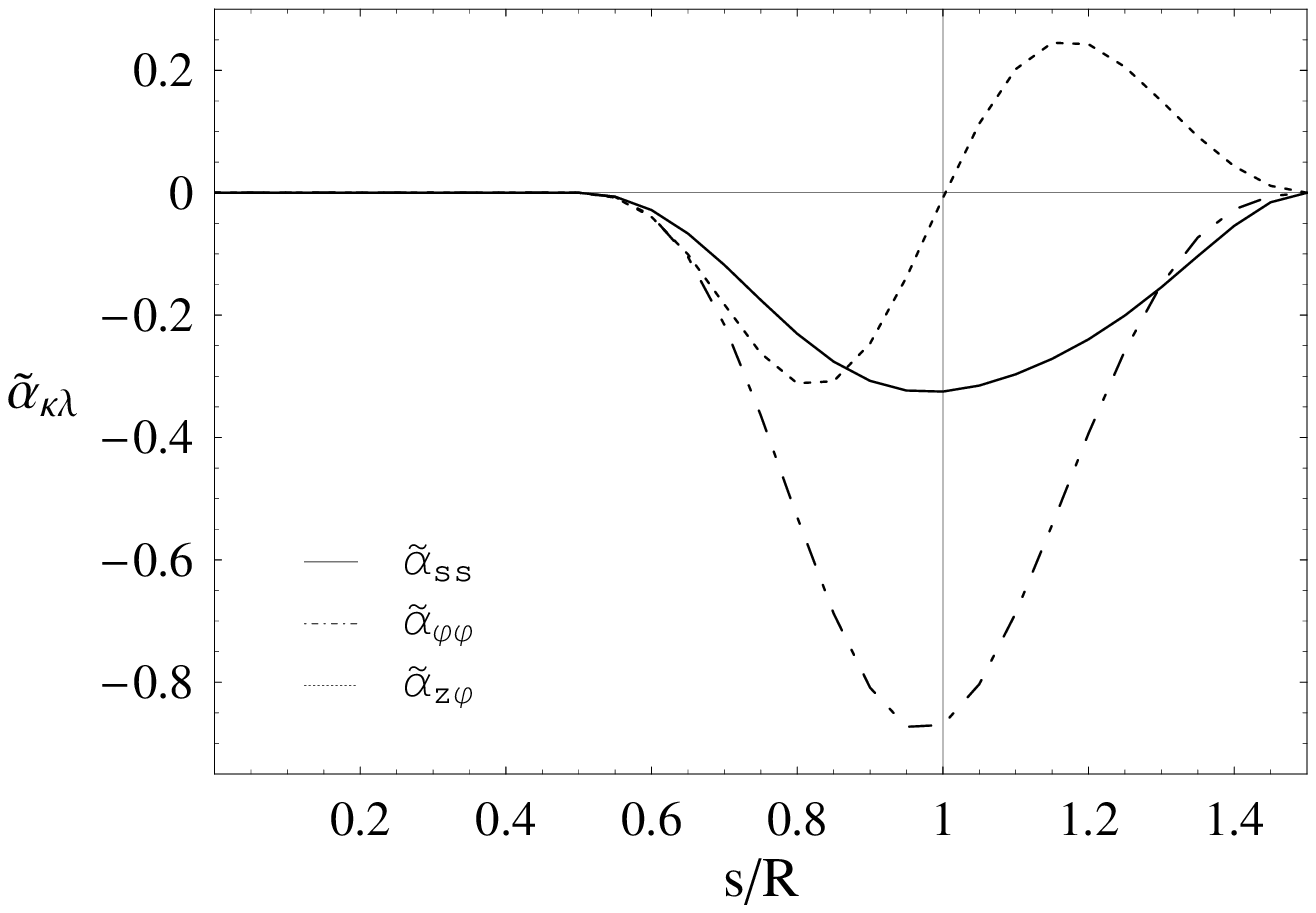}} $\;\;$\subfigure[]{\includegraphics[%
  clip,
  width=7.3cm,
  keepaspectratio]{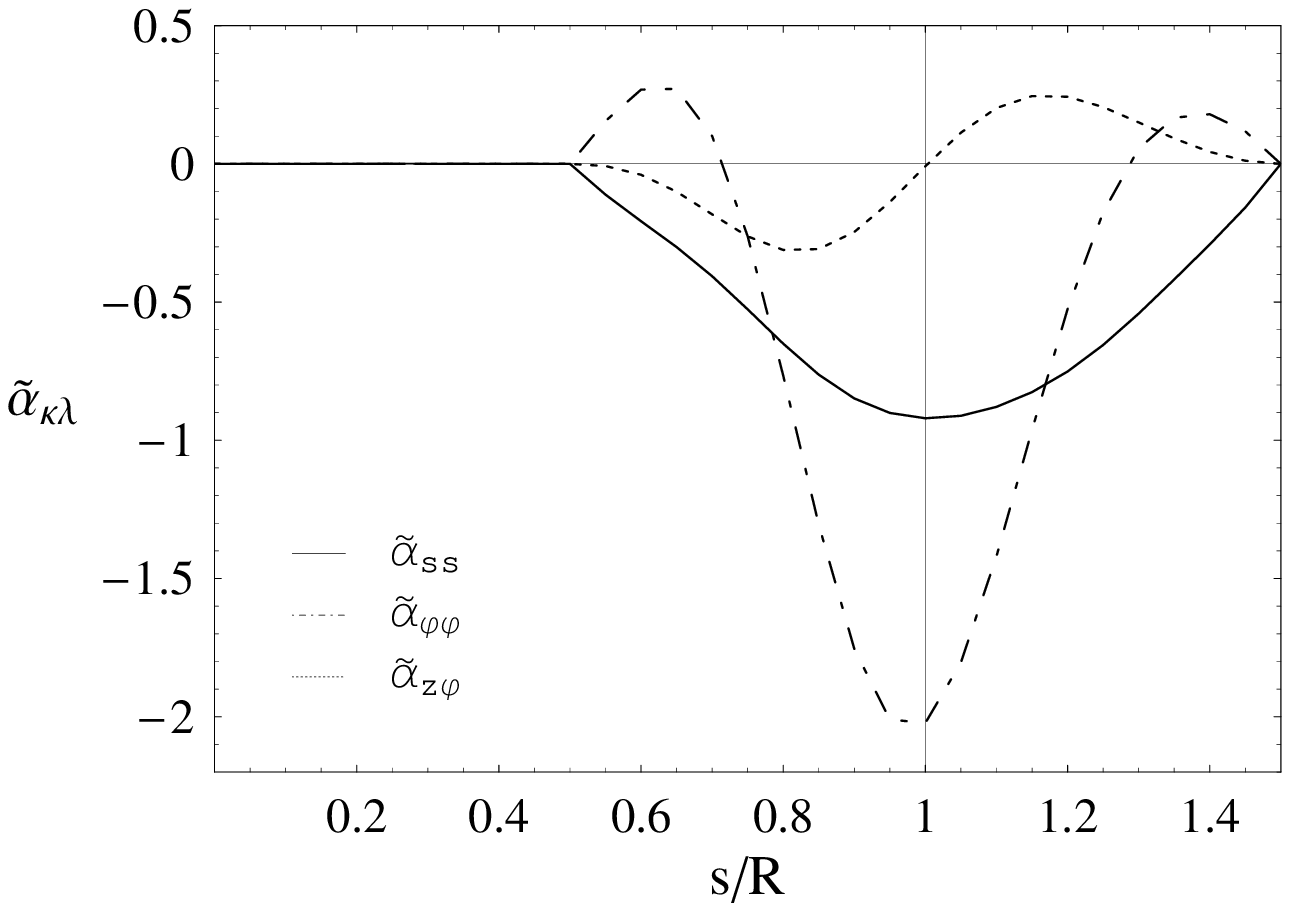}}\end{center}

\caption{Radial profile of the dimensionless quantity $\widetilde{\alpha}_{\kappa\lambda}$
for (a) FW1 and (b) FW2, both for $n=4$ and $\delta=0.5$.}
\end{figure}

\section*{Appendix C: Numerical approach}

The correct handling of the non-local boundary conditions for the
magnetic field is a notorious problem for the simulation of dynamos
in non-spherical domains. Here, the kinematic eigenvalue problem in
finite cylinders is solved by the integral equation approach (Stefani
\textit{et al.} 2000, Xu \textit{et al.} 2004a, Xu \textit{et al.}
2004b, Xu \textit{et al.} 2006). Basically, we use the following three
integral equations: \begin{eqnarray*}
\bB({\mathbf{r}}) & = & \frac{\mu\sigma}{4\pi}\int_{V}\frac{(\alpha\circ\bB({\mathbf{r}}'))\times{(\mathbf{r}-\mathbf{r}')}}{|{\mathbf{r}}-{\mathbf{r}}'|^{3}}{\rm d}V'-\frac{\mu\sigma p}{4\pi}\int_{V}\frac{{\mathbf{A}}({\mathbf{r}}')\times({\mathbf{r}}-{\mathbf{r}}')}{|{\mathbf{r}}-{\mathbf{r}}'|^{3}}{\rm d}V'\\
\qquad\qquad\qquad &  & -\frac{\mu\sigma}{4\pi}\int_{S}\phi({\bzeta}')\:{\mathbf{n}}({\bzeta}')\times\frac{{\mathbf{r}}-{\bzeta}'}{|{\mathbf{r}}-{\bzeta}'|^{3}}{\rm d}S',\qquad\qquad\qquad\qquad\qquad\mathrm{(C.1)}\end{eqnarray*}
\begin{eqnarray*}
\frac{1}{2}\phi({\bzeta}) & = & \frac{1}{4\pi}\int_{V}\frac{(\alpha\circ\bB({\mathbf{r}}'))\cdot({\bzeta}-{\mathbf{r}}')}{|{\bzeta}-{\mathbf{r}}'|^{3}}{\rm d}V'-\frac{p}{4\pi}\int_{V}\frac{{\mathbf{A}}({\mathbf{r}}')\cdot({\bzeta}-{\mathbf{r}}')}{|{\bzeta}-{\mathbf{r}}'|^{3}}{\rm d}V'\\
\qquad\qquad\qquad &  & -\frac{1}{4\pi}\int_{S}\phi({\bzeta}')\:{\mathbf{n}}({\bzeta}')\cdot\frac{{\bzeta}-{\bzeta}'}{|{\bzeta}-{\bzeta}'|^{3}}{\rm d}S',\qquad\qquad\qquad\quad\;\;\quad\quad\quad\mathrm{(C.2)}\end{eqnarray*}
\begin{eqnarray*}
\qquad\qquad{\mathbf{A}}({\mathbf{r}}) & = & \frac{1}{4\pi}\int_{V}\frac{\bB({\mathbf{r}}')\times({\mathbf{r}}-{\mathbf{r}}')}{|{\mathbf{r}}-{\mathbf{r}}'|^{3}}{\rm d}V'+\frac{1}{4\pi}\int_{S}{\mathbf{n}}({\bzeta}')\times\frac{\bB({\bzeta}')}{|{\mathbf{r}}-{\bzeta}'|}{\rm d}S',\quad{\rm (C.3)}\end{eqnarray*}
 where $\bB$ is the magnetic field, ${\mathbf{A}}$ the vector potential,
$\phi$ the electric potential, ${\mathbf{n}}$ the outward directed
unit vector at the boundary $S$. The complex constant $p$ contains
as its real part the growth rate and as its the imaginary part the
frequency of the eigenfield. The matrix $\alpha$ represents the $\alpha$-effect
defined by (\ref{eq:matrix-alpha}) or by (\ref{eq:matrix-alpha-z}).

The reduction of the problem to cylindrical problems with azimuthal
waves $\exp\left({\textrm{i}}\, m\varphi\right)$ was described in
Xu \textit{et al.} (2006). Finally, we end up with a generalised eigenvalue
problem for the critical dynamo number $C_{\alpha}^{c}$ (in the steady
case), or for the complex constant $p$ (in the unsteady case). The
QR method is employed to solve this eigenvalue problem which gives
also the eigenmodes of the magnetic field.

\section*{References}

\noindent Aubert, J. and Wicht, J., Axial versus equatorial dipolar
dynamo models with implications for planetary magnetic fields. \textit{Earth.
Plan. Sci. Lett.}, 2004, \textbf{221}, 409-419. \medskip{}

\noindent Avalos-Z\'{u}\~{n}iga, R., Plunian, F., and R\"{a}dler, K.-H.,
Rossby waves and $\alpha$-effect. 2007, to be submitted. \medskip{}

\noindent Busse, F.H., Model of geodynamo. \textit{Geophys. J. R.
Astron. Soc.}, 1975, \textbf{42}, 437-459. \medskip{}

\noindent Gailitis, A., Self-excitation conditions for a laboratory
model of geomagnetic dynamo. \textit{Magnetohydrodynamics}, 1967,
\textbf{3}, 23-29. \medskip{}

\noindent Giesecke, A., R\"{u}diger, G. and Elstner, D., Oscillating
$\alpha^{2}$-dynamos and the reversal phenomenon of the global geodynamo.
\textit{Astron. Nachr.}, 2005a, \textbf{326}, 693-700. \medskip{}

\noindent Giesecke, A., Ziegler, U. and R\"{u}diger, G., Geodynamo
$\alpha$-effect derived from box simulations of rotating magnetoconvection.
\textit{Phys. Earth Planet. Inter.}, 2005b, \textbf{152}, 90-102.

\noindent \medskip{} Grote, E. and Busse, F.H., Hemispherical dynamos
generated by convection in rotating spherical shells. \textit{Phys.
Rev. E}, 2000, \textbf{62}, 4457-4460. \medskip{}

\noindent Gubbins, D., Barber, C.N., Gibbons, S. and Love, J.J., Kinematic
dynamo action in a sphere. II Symmetry selection. \textit{Proc. R.
Soc. Lond. A}, 2000, \textbf{456}, 1669-1683. \medskip{}

\noindent Ishihara, N. and Kida, S., Dynamo mechanism in a rotating
spherical shell: competition between magnetic field and convection
vortices. \textit{J. Fluid Mech.}, 2002, \textbf{465}, 1-32. \medskip{}

\noindent Melbourne, I., Proctor, M.R.E., \& Rucklidge, A.M., A heteroclinic
model of geodynamo reversals and excursions, in: \textit{Dynamo and
Dynamics, a Mathematical Challenge} (eds. P. Chossat, D. Armbruster
and I. Oprea), Kluwer, Dordrecht, 2001, pp. 363-370. \medskip{}

\noindent Olson, P., Christensen, U. and Glatzmaier, G.A., Numerical
modelling of the geodynamo: Mechanisms of field generation and equilibration.
\textit{J. Geophys. Res.}, 1999, \textbf{104}, 10383-10404. \medskip{}

\noindent Phillips, C.G., Mean dynamos, 1993, Sydney University Ph.D.
Thesis

\noindent R\"{a}dler, K.-H., Some new results on the generation of
magnetic fields by dynamo action. \textit{Mem. Soc. Roy. Sci. Liege,
Ser. 6}, 1975, \textbf{VIII}, 109-116. \medskip{}

\noindent R\"{a}dler, K.-H., Mean-Field Approach to Spherical Dynamo
Models, \textit{Astron. Nachr.}, 1980, \textbf{301}, 101-129. \medskip{}

\noindent R\"{a}dler, K.-H, Investigations of spherical kinematic
mean-field dynamos. \textit{Astron. Nachr.}, 1986, \textbf{307}, 89-113.
\medskip{}

\noindent R\"{a}dler, K.-H., Apstein, E., Rheinhardt, M. and Sch\"{u}ler,
M., The Karlsruhe dynamo experiment. A mean field approach, \textit{Stud.
Geophys. Geodaet.}, 1998, \textbf{42}, 224-231. \medskip{}

\noindent R\"{a}dler, K.-H.,Rheinhardt,M., Apstein, E. and Fuchs,
H., On the mean-field theory of the Karlsruhe dynamo experiment. \textit{Nonlin.
Proc. Geophys.}, 2002, \textbf{9}, 171-187. \medskip{}

\noindent R\"{u}diger, G., Rapidly rotating $\alpha^{2}$-dynamos
models. \textit{Astron. Nachr.}, 1980, \textbf{301}, 181-187. \medskip{}

\noindent R\"{u}diger, G. and Elstner, D., Non-axisymmetry vs. axi-symmetry
in dynamo-excited stellar magnetic fields. \textit{Astron. Astrophys.},
1994, \textbf{281}, 46-50. \medskip{}

\noindent R\"{u}diger, G., Elstner, D. and Ossendrijver M., Do spherical
$\alpha^{2}$-dynamos oscillate? \textit{Astron. Astrophys.}, 2003,
\textbf{406}, 15-21. \medskip{}

\noindent Sarson, G.R . and Jones, C.A., A convection driven geodynamo
reversal model. \textit{Phys. Earth Planet. Inter.}, 1999, \textbf{111},
3-20. \medskip{}

\noindent Schaeffer N. and Cardin, P., Quasi-geostrophic kinematic
dynamos at low magnetic Prandtl number. \textit{Earth Planet. Sci.
Lett.}, 2006, \textbf{245}, 595-604. \medskip{}

\noindent Stefani, F., Gerbeth, G. and R\"{a}dler, K.-H., Steady
dynamos in finite domains: an integral equation approach. \textit{Astron.
Nachr.}, 2000, \textbf{321}, 65-73. \medskip{}

\noindent Stefani, F. and Gerbeth, G., Asymmetry polarity reversals,
bimodal field distribution, and coherence resonance in a spherically
symmetric mean-field dynamo model. \textit{Phys. Rev. Lett.}, 2005,
\textbf{94}, Art. No. 184506. \medskip{}

\noindent Stefani, F., Gerbeth, G., G\"{u}nther, U. and Xu, M., Why
dynamos are prone to reversals. \textit{Earth Planet. Sci. Lett.},
2006a, \textbf{143}, 828-840. \medskip{}

\noindent Stefani, F., Gerbeth, G. and G\"{u}nther, U., A paradigmatic
model of Earth's magnetic field reversals. \textit{Magnetohydrodynamics},
2006b, \textbf{42}, 123-130. \medskip{}

\noindent Stefani, F., Xu, M., Gerbeth, G., Ravelet, F., Chiffaudel,
A., Daviaud, F. and Leorat, J., Ambivalent effects of added layers
on steady kinematic dynamos in cylindrical geometry: application to
the VKS experiment. \textit{Eur. J. Mech. B/Fluids}, 2006c, \textbf{25},
894-908. \medskip{}

\noindent Stieglitz R. and M\"{u}ller U., Experimental demonstration
of the homogeneous two-scale dynamo. \textit{Phys. Fluids}, 2001,
\textbf{13}, 561-564. \medskip{}

\noindent Tilgner, A., Small scale kinematic dynamos: beyond the $\alpha$-effect,
\textit{Geophys. Astrophys. Fluid Dyn.}, 2004, \textbf{98}, 225-234.
\medskip{}

\noindent Weisshaar, E, A numerical study of $\alpha^{2}$- dynamos
with anisotropic $\alpha$-effect. \textit{Geophys. Astrophys. Fluid
Dyn.}, 1982, \textbf{21}, 285-301. \medskip{}

\noindent Wicht, J. and Olson, P., A detailed study of the polarity
reversal mechanism in a numerical dynamo model. \textit{Geochem. Geophys.
Geosys.}, 2004, \textbf{5}, Art. No Art. No. Q03H10. \medskip{}

\noindent Xu, M., Stefani, F. and Gerbeth, G., The integral equation
method for a steady kinematic dynamo problem. \textit{J. Comp. Phys.},
2004a, \textbf{196}, 102-125. \medskip{}

\noindent Xu, M., Stefani, F. and Gerbeth, G. Integral equation approach
to time-dependent kinematic dynamos in finite domains. \textit{Phys.
Rev. E}, 2004b, \textbf{70}, Art. No. 056305. \medskip{}

\noindent Xu, M., Stefani, F. and Gerbeth, G., The integral equation
approach to kinematic dynamo theory and its application to dynamo
experiments in cylindrical geometry. in: \textit{Proceedings of ECCOMAS
CFD 2006}, (eds: P. Wesseling, E. Onate, J. Periaux), TU Delft, paper
497 (CD). \medskip{}

\noindent Yoshimura, H., Wang, Z. and Wu, F., Linear astrophysical
dynamos in rotating spheres: mode transition between steady and oscillatory
dynamos as a function of dynamo strength and anisotropic turbulent
diffusivity. \textit{Astrophys. J.}, 1984, \textbf{283}, 870-878.\newpage

\medskip{}
\end{document}